\documentclass[a4paper,11pt]{article}

\usepackage{jheppub} 

\usepackage{amsmath}
\usepackage{rotating}
\usepackage{bm}
\usepackage{amssymb}
\usepackage{graphicx}

\newcommand{\Op}{\mathcal{O}}
\newcommand{\HS}{S}
\newcommand{\HBS}{\mathbb S}

\newcommand{\M}{M}

\newcommand{\barom}{\bar\omega}

\newcommand{\signN}{(-1)^r}

\def\z#1{{{\zeta_#1}}}
\def\zz#1#2{\zeta_{#1#2}}

\newcommand{\cN}{{\cal N}}
\newcommand{\Z}{{\mathcal Z}}
\newcommand{\X}{{\mathcal X}}
\newcommand{\Y}{{\mathcal Y}}
\newcommand{\cD}{\mathcal{D}}
\newcommand{\w}{\omega}

\newcommand{\zzeta}{\bm{{{\zeta}}}}

\newcommand{\tx}{\times}

\def\z#1{{{\zeta_#1}}}

\title{\boldmath Exact result in $\cN=4$ SYM theory: \\[1mm] Generalised double-logarithmic equation}

\author[a]{V.~N.~Velizhanin}

\affiliation[a]{Theoretical Physics Division\\
NRC ``Kurchatov Institute''\\
Petersburg Nuclear Physics Institute\\
Orlova Roscha, Gatchina\\
188300 St.~Petersburg, Russia}

\emailAdd{velizh@thd.pnpi.spb.ru}

\abstract{
We present the new results for the generalised double-logarithmic equation, obtained from the analytical continuation of the seven-loop anomalous dimension of twist-2 operators in the planar $\cN=4$ SYM theory. 
The double-logarithmic equation is related to the special asymptotic of the scattering amplitudes, when the large logarithms of the energy of scattering particles are appeared and should be summed in all order of perturbative theory. These large logarithms correspond to the poles of the analytically continued anomalous dimension. The generalised double-logarithmic equation includes the subleading logarithms.
We have found, that the expansion of the generalised double-logarithmic equation can be ressumed in the form of rational functions with simple denominator.
The solution of the generalised double-logarithmic equation provides a lot of information about the poles of the analytically continued anomalous dimension in all orders of perturbative theory.
We have found also the generalised double-logarithmic equation for the analytically continued anomalous dimension near the value, which is related with BFKL-equation.
}

\begin{document} 
\maketitle
\flushbottom

\section{Introduction}

Great progress has been made in the study of the composite operators in the maximally extended $\cN=4$ Supersymmetric Yang-Mills (SYM) theory and their pairs string states in AdS space through AdS/CFT-correspondence~\cite{Maldacena:1997re,Witten:1998qj,Gubser:1998bc}. The most studied operators in the $\cN=4$ SYM theory consist of scalar fields.
The simplest operator is build from one complex scalar field, usually denoting as $\Z$
\begin{equation}
\Op_{\Z^J}={\mathrm{tr}\,}\Z\Z\Z\cdots\Z\equiv{\mathrm{tr}\,}\Z^J
\end{equation}
and it is protected from the quantum corrections.
To construct other operators with minimal efforts one can either add other fields of $\cN=4$ SYM theory, for example, a number of another complex scalar fields, usually denoting as $\X$ or $\Y$
\begin{eqnarray}
\Op_{\Z^J\X}&=&
{\mathrm{tr}\,}\Z^{J-k}\X\Z^k,\nonumber\\
\Op_{\Z^J\X^2}&=&
{\mathrm{tr}\,}\Z^{J-k}\X\X\Z^k,\nonumber\\
\Op_{\Z^J\X^\M}&=&
{\mathrm{tr}\,}\Z^{J-k}\X^\M\Z^k,\ldots\label{OpZZX}
\end{eqnarray}
either add a number of covariant derivatives $\cD_\mu$ between the fields. In the last case, taking only two fields and $\M$ covariant derivatives $\cD_{\mu_i}$ we obtain twist-2 operators with the Lorentz spin~$\M$
\begin{equation}
\Op_{\mathrm{twist-2}}={\mathrm{tr}\,}\Z\cD_{\mu_1}\cD_{\mu_2}\cdots\cD_{\mu_\M}\Z\equiv \Z\cD_\mu^\M\Z\,,\label{OpM}
\end{equation}
which are the well-know from the study of a Deep-Inelastic Scattering (DIS), where they are appeared under an operator product expansion.
Starting from \mbox{$\M=2$} operators~(\ref{OpM}) are unprotected and receive the quantum corrections to the canonical dimension of the operator
\begin{equation}
\Delta={\mathrm{Canonical~dimension}}+\gamma
\end{equation}
where $\gamma$ is the anomalous dimension, generating by the quantum corrections.
There are a lot of reasons to obtain the result for the anomalous dimension of twist-2 operators~(\ref{OpM}) with arbitrary Lorentz spin $\M$ in $\cN=4$ SYM theory. Initially, such results were very interesting for the study of AdS/CFT-correspondence. In principle, such task can be performed by the direct diagrammatic calculations in the first several order of perturbative theory. In the leading order the computations are rather trivial~\cite{Anselmi:1998ms,Lipatov:2001fs,Arutyunov:2001mh,Dolan:2001tt} and the general result for arbitrary Lorentz spin $\M$ has the following form~\cite{Lipatov:2001fs,Dolan:2001tt,Kotikov:2002ab}:
\begin{equation}
\gamma^{(2)}(\M)=\sum_{i=1}^\M\frac{1}{i}\ ,\qquad\qquad
\gamma(\M)=
\sum_{\ell=1}g^{2\ell}\,\gamma^{(2\ell)}(\M)
\end{equation}
During the next-to-leading order calculations it was found~\cite{Kotikov:2003fb}, that the anomalous dimensions have a lot of remarkable properties in particularly confirming the maximal transcendentality principle~\cite{Kotikov:2002ab}, which relates the results in Quantum Chromodynamics (QCD) and in $\cN=4$ SYM theory and roughly states, that the most complicated part of the corresponding result in QCD gives a desired result in $\cN=4$ SYM theory.

Later it was found, that there is a more powerful and more general way to compute not only the anomalous dimensions of the composite operators, but also the energies of their pairs string states. This was done through an identification of composite operators with a spin chain~\cite{Minahan:2002ve}, so the calculation of the anomalous dimension was reduced to the computation of the energy of the spin chain state. The last problem can be solved with the help of so called integrability, for example, using the Bethe ansatz method~\cite{Bethe:1931hc,Faddeev:1981ft,Reshetikhin:1983vw}.

In this way, a number of exact results were obtained for the unprotected operators, among which the Asymptotic Bethe Ansatz (ABA)~\cite{Beisert:2004hm,Beisert:2005fw,Beisert:2006ez} for the operators/strings with large quantum numbers (large $J$), Thermodynamic Bethe Ansatz (TBA)~\cite{Bombardelli:2009ns,Gromov:2009bc,Arutyunov:2009ur}, \mbox{Y-system}~\cite{Gromov:2009tv,Cavaglia:2010nm} and Quantum Spectral Curve (QSC)~\cite{Gromov:2013pga,Gromov:2014caa} for the arbitrary composite operators, at least in $sl(2)$ sector (see Ref.~\cite{Beisert:2010jr} for review)\footnote{Earlier, similar integrability was discovered in QCD in the Regge limit~\cite{Lipatov:1993qn,Lipatov:1993yb,Faddeev:1994zg} and for some operators~\cite{Braun:1998id}.}. 
However, all these results are concerned the operators with some fixed integer value of $\M$, which is the number of covariant derivatives in twist-2 operator or the number of impurities in the integrable language and for some limiting cases (the most important is large $\M$ limit).
For the applicability in the realistic models such as QCD the most interesting result is related with the so called \mbox{small-$x$} physics, which corresponds to the special limit of the general expression for the splitting function related with the anomalous dimension through a Mellin transformation. In this limit the large logarithms $\ln x$ appear, which, in principle, should be summed in all order of perturbative theory. Such resummation can be performed in some cases, for example, with the Balitsky-Fadin-Kuraev-Lipatov (BFKL) equation~\cite{Lipatov:1976zz,Kuraev:1977fs,Balitsky:1978ic,Fadin:1998py} or with the double-logarithmic equation~\cite{Gorshkov:1966qd,Kirschner:1983di} depending on the process under consideration. Both these equations originate from the study of the scattering amplitudes in the high-energy asymptotic, which are expanded over the powers of $\ln x$ for the small-$x$ limit. For the BFKL-equation the QSC-approach allow to compute pomeron eigenvalue~\cite{Alfimov:2014bwa,Gromov:2015vua,Alfimov:2018cms}, while the double-logarithmic equation is not understanded very well from the relation with integrability.

In this paper we will consider the double-logarithmic equation for the twist-2 operators in the planar $\cN=4$ SYM theory and its generalisation to the subleading logarithm approximations using the results for the seven-loop anomalous dimension of twist-2 operators in planar $\cN=4$ SYM theory~\cite{Kotikov:2002ab,Kotikov:2003fb,Kotikov:2004er,Staudacher:2004tk,Kotikov:2007cy,Bajnok:2008qj,Lukowski:2009ce,Marboe:2014sya,Marboe:2016igj} and a database for the analytical continuation for the relevant function~\cite{Velizhanin:2020avm}.
In Section~\ref{Section:1} we review the alternative representation for the anomalous dimension of twist-2 operators and its relations with the exact results, coming from the generalised double-logarithmic equation.
Section~\ref{Section:2} gives a short introduction to the derivation of the double-logaritmic equation from the direct diagrammatic computations and from the infra-red evolution equation.
In Section~\ref{Section:3} we introduce the generalised double-logarithmic equation, providing the detailed information about its origin, its properties and its solution. Moreover, we present the similar generalised double-logarithmic equation, obtained from the analytical continuation of the anomalous dimension near different value. 
In Appendix~\ref{App:ReconAD} we show, how the anomalous dimension can be reconstructed from the information, provided by the generalised double-logarithmic equation.
Appendix~\ref{Section:DLgenerALL} and Appendix~\ref{DLoddF} contain the expansion of the right-hand side of the generalised double-logarithmic equation up to seven loops and transcendental number with weight $12$ ($\zz12$ and similar).

\section{Analytical properties of anomalous dimension}\label{Section:1}

Twist-2 operators are exceptional among all others as their anomalous dimension has a lot of remarkable properties.
As well known, the anomalous dimension of twist-2 operators is expressed through the nested harmonic sums defined as (see \cite{Vermaseren:1998uu}):
\begin{equation}\label{vhs}
\HS_a (M)=\sum^{M}_{j=1} \frac{(\mbox{sgn}(a))^{j}}{j^{\vert a\vert}}\, , \qquad
\HS_{a_1,\ldots,a_n}(M)=\sum^{M}_{j=1} \frac{(\mbox{sgn}(a_1))^{j}}{j^{\vert a_1\vert}}
\,\HS_{a_2,\ldots,a_n}(j)\, ,
\end{equation}
where $M$ is a positive integer number.
In the leading order of perturbative theory, the result for the anomalous dimension of twist-2 operators~(\ref{OpM}) in $\cN=4$ SYM theory is given by~\cite{Lipatov:2001fs,Dolan:2001tt,Kotikov:2002ab}:
\begin{equation}\label{LOAD}
\gamma^{\mathrm{LO}}(M)=4\HS_1(M)\,.
\end{equation}
The simplest harmonic sum $\HS_1$ can be written in the following way for all values of its argument (analytically continued to the complex plane):
\begin{equation}\label{S1AC}
\HS_1(M)=\sum^{M}_{j=1}\frac{1}{j}=\sum^{\infty}_{j=1}\frac{1}{j}-\sum^{\infty}_{j=M+1}\frac{1}{j}=\sum^{\infty}_{j=1}\frac{1}{j}-\sum^{\infty}_{j=1}\frac{1}{j+M}\ ,
\end{equation}
i.e. it has the simple poles for all negative values of $\M$.
Only one meromorphic function has the simple poles with residues $(-1)$ in all negative integer values of its argument and this is the polygamma-function $\Psi(\M)$:
\begin{equation}\label{S1}
\Psi(M)=\sum_{j=1}^{\infty}\left(\frac{1}{j}-\frac{1}{j+M-1}\right)-\gamma_E\overset{M \in\, \mathbb{Z}^{{}^{+}}}{=}S_1(M-1)+\Psi(1)\,.
\end{equation}
All other nested harmonic sums being produced from the simplest harmonic sum, which is the polygamma-function, have the same properties: they can be uniquely identified from the corresponding expressions, which include their residues near the negative integer values. Such expressions can be extracted from our database~\cite{Velizhanin:2020avm}, which was obtained by means of the inverse Mellin transformation from the harmonic sums to the harmonic polylogarithms in $x$-space~\cite{Vermaseren:1998uu,Remiddi:1999ew} and extraction of the small-$x$ logarithms, which correspond to the poles in the initial $\M$-space.
For example, the simplest non-trivial nested harmonic sum with alternating summation has the following pole structure, being analytically continued from even integer $\M$ near $\M=-r,\ r=1,2,3,\ldots$ 
\begin{equation}\label{Sm21residues}
\check{S}_{-2,1}(r)=
\frac{1}{\omega^3}\times 0
+\frac{1}{\omega^2}\times 0
+\frac{(-1)^{r}}{\omega}\Big[S_{-2}(r-1)-\z2\Big]\,,\qquad \mathrm{vs.}\ \
\check{S}_{1}(r)=-\frac{1}{\omega}\,.
\end{equation}
Replaced $\omega\to j+M$ and $r\to j$ we obtain, by analogy with Eq.~(\ref{S1}), the dispersion (Mittag-Leffler) representation\footnote{By analogy with the represetation for the Baxter function in Ref.~\cite{DeVega:2001pu}.}\footnote{Similar representation for the analytically continued harmonic sums was used in Refs.~\cite{Fadin:1998py,Kotikov:2000pm,Velizhanin:2015xsa} and in unpublished work of L.~Lipatov and A.~Onishchenko (2004).} for $S_{-2,1}(\M)$
\begin{equation}\label{Sm21}
\hat{S}_{-2,1}(\M)=
\sum_{j=1}^{\infty}\frac{(-1)^{j}}{j+\M}\Big[S_{-2}(j-1)-\z2\Big]
+S_{-2,1}^{\infty},\qquad
S_{-2,1}^{\infty}=S_{-2,1}({\infty})=
-\frac{5}{8}\z3\,,
\end{equation}
where $\M$ is now the arbitrary complex number. For the analytic continuation from odd integers $\M$ we should replace $(-1)^r$ to $-(-1)^r$ in Eq.~(\ref{Sm21residues}) and $(-1)^j$ to $-(-1)^j$ in Eq.~(\ref{Sm21}).

In fact, the above dispersion representation is related to the expression obtained with the hep of the usual analytic continuation of the nested harmonic sums~\cite{Gonzalez-Arroyo:1979guc,Yndurain:2006amm,Kotikov:2005gr,Ablinger:2013cf}, 
which in the case of the considered harmonic sum $S_{-2,1}$ looks like
\begin{eqnarray}
S_{-2,1}(\M)&\to&
\sum_{j=1}^{\infty}
\sum_{k=1}^{\infty}\Bigg[
\frac{(-1)^k}{(k+\M)^2}\frac{1}{j+k+\M}
-\frac{(-1)^k}{(k+\M)^2}\frac{1}{j}
-\frac{(-1)^k}{k^2}\frac{1}{j+k}
+\frac{(-1)^k}{k^2}\frac{1}{j}\Bigg]\qquad\label{UsualACSm21}
\\&=&
\sum_{j=1}^{\infty}
\sum_{k=1}^{\infty}
\frac{(-1)^k}{(k+\M)^2}\frac{1}{j+k+\M}
-S_1^{\infty}\sum_{k=1}^{\infty}\frac{(-1)^k}{(k+\M)^2}
-\zeta_{1,-2}
-\frac{\z2}{2} S_1^{\infty}\label{UsualACSm21S},
\end{eqnarray}
where $\zeta_{1,-2}=-\z2/2\,S_1^\infty+5/8\,\z3$.
Using the decomposition of the first term in the form
\begin{equation}
\frac{1}{(k+\M)^2}\frac{1}{j+k+\M}=-\frac{1}{j^2 (k+\M)}+\frac{1}{j^2 (j+k+\M)}+\frac{1}{j (k+\M)^2}
\end{equation}
it can be written as
\begin{eqnarray}\label{rewritefirst}
\sum_{j=1}^{\infty}
\sum_{k=1}^{\infty}
\frac{1}{j^2}\frac{(-1)^k}{j+k+\M}
-\z2\sum_{k=1}^{\infty}\frac{(-1)^k}{(k+\M)}
+S_1^{\infty}\sum_{k=1}^{\infty}\frac{(-1)^k}{(k+\M)^2}
\end{eqnarray}
Then, we change $j+k\to j$ for the first term and obtain 
\begin{equation}\label{resumfirst}
\sum_{j=1}^{\infty}
\sum_{k=1}^{\infty}
\frac{1}{j^2}\frac{(-1)^k}{j+k+\M}=
\sum_{j=2}^{\infty}
\frac{1}{j+\M}
\sum_{k=1}^{j-1}
\frac{(-1)^k}{(j-k)^2}\,,
\end{equation}
where
\begin{equation}\label{resummin}
\sum_{k=1}^{j-1}
\frac{(-1)^k}{(j-k)^2}=
(-1)^jS_{-2}(j-1)\,.
\end{equation}
Substituting Eqs.~(\ref{resummin}), (\ref{resumfirst}) and (\ref{rewritefirst}) into Eq.~(\ref{UsualACSm21S}) we obtain
\begin{equation}
S_{-2,1}(\M)\to
\sum_{j=1}^{\infty}
\frac{(-1)^j}{(j+\M)^2}\Big(\HS_{-2}(j-1)-\z2\Big)
-\frac{5}{8}\z3\,,
\end{equation}
which is the same expression as Eq.~(\ref{Sm21}). One can check numerically, that both representations Eq.~(\ref{Sm21}) and Eq.~(\ref{UsualACSm21}) give the same result for the arbitrary complex values of $\M$.
Thus, the analytically continued harmonic sum $S_{-2,1}(M)$ in Eq.~(\ref{UsualACSm21S}) can be rewritten in the form of dispersion  representation (\ref{Sm21}), which was obtained from the expansion near the negative integer values~(\ref{Sm21residues}). Similar transformations can be performed for other nested harmonic sums.

Moreover, it is easy to see, that the number of the harmonic sums in the expression for the anomalous dimension of twist-2 operators at $\ell$-order of the perturbative theory, which is equal to $((1 - \sqrt{2})^k + (1 + \sqrt{2})^k)/2$ for $k=2\ell-1$, is the same, as the number of the unique pole structures for the expansion near the negative integer value of the analytically continued harmonic sums. For example, at two-loop order the basis from the usual harmonic sums contains the following seven sums with transcendentality $3$:
\begin{equation}
\left\{S_{-3},S_3,S_{2,1},S_{1,2},S_{-2,1},S_{1,-2},S_{1,1,1}\right\}\,,
\end{equation}
while the two-loop result for positive $M$ is the following~\cite{Kotikov:2003fb}:
\begin{equation}\label{NLOAD}
\gamma^{\mathrm{NLO}}(M) =8\Big[\,2\,S_{2,1}+2\,S_{1,2}-S_3-S_{-3}+2\,S_{1,-2}\,\Big]\,.
\end{equation}
The analytic continuation of the harmonic sums near negative integer values has the following form
\begin{eqnarray}&&
\Bigg\{
-\frac{\signN}{\w^3},\
-\frac{1}{\w^3},\
\frac{S_2-{\z2}}{\w},\
-\frac{S_1}{\w^2}-\frac{S_2}{\w},\
\frac{{\signN}\big(S_{-2}-{\z2}\big)}{\w},\nonumber\\&&\hspace{10mm}
\frac{\frac{{\z2}}{2}-{\signN} \big(S_2+ \frac{{\z2}}{2}\big)}{\w}-\frac{{\signN}S_1}{\w^2},\
\frac{S_2-S_{1,1}}{\w}\Bigg\},\qquad\qquad S_{\vec{a}}=S_{\vec{a}}(r-2)\qquad
\end{eqnarray}
and the unique structures from the above pole expressions are
\begin{eqnarray}&&
\Bigg\{
\frac{\signN}{\w^3},\
\frac{1}{\w^3},\
\frac{S_2}{\w},\
\frac{S_1}{\w^2},\
\frac{{\signN}S_{-2}}{\w},\
\frac{{\signN}S_1}{\w^2},\
\frac{S_{1,1}}{\w}\Bigg\}\,.
\end{eqnarray}
Other unique pole structures come form the harmonic sums with index $(-1)$, which do not enter into the expressions for the anomalous dimensions of twist-2 operators:
\begin{eqnarray}
\Big\{S_{-1,-2},S_{-1,2},S_{-2,-1},S_{2,-1}\Big\}&\to&
\Bigg\{
\frac{-S_{-2}+\frac{{\signN} {\z2}}{2}-\frac{{\z2}}{2}}{\w}-\frac{S_{-1}}{\w^2},
\nonumber\\&&\hspace*{-10mm}
-\frac{{\signN} S_{-1}}{\w^2}-\frac{{\signN} S_{-2}}{\w},
\frac{S_{-2}+\frac{{\z2}}{2}}{\w}+\frac{{\signN}{\ln2}-{\ln2}}{\w^2},
\nonumber\\&&\hspace*{-10mm}
\frac{{\signN} \big(S_2+\frac{{\z2}}{2}\big)}{\w}+\frac{{\ln2}- {\signN}{\ln2}}{\w^2}
\Bigg\}
\end{eqnarray}
with the following unique pole structures
\begin{equation}
\Bigg\{
\frac{S_{-1}}{\w^2},\
\frac{{\signN} S_{-1}}{\w^2},\ 
\frac{S_{-2}}{\w},\ 
\frac{{\signN}S_2}{\w}
\Bigg\}
\end{equation}
and $S_{\pm 1,\pm 1,\pm 1}\to(\pm 1)^r S_{\pm 1,\pm 1}/\w$. The presence in the pole expression for the anomalous dimension, for example, $(-1)^rS_{-2}/\omega$ means that the expression for the anomalous dimension through the usual harmonic sums will contain $S_{-2,1}$ and so on.

Thus, if we know the pole structure for the anomalous dimension, we can write the dispersion representation for it.
Up to two-loop order we have from our database~\cite{Velizhanin:2020avm} for Eq.~(\ref{LOAD}) and~Eq.~(\ref{NLOAD})
\begin{eqnarray}
\breve{\gamma}^{\mathrm{LO}}(r) &=&
4\bigg[-\frac{1}{\w}
\bigg],
\label{LOADpole}\\
\breve{\gamma}^{\mathrm{NLO}}(r) &=&
8
\bigg[
\frac{(1+(-1)^{r})}{\omega ^{3}}
-2\frac{(1+(-1)^{r})}{\omega ^{2}} S_{1}(r-1) \nonumber \\
&&\qquad-\Big((1+(-1)^{r})\zeta (2)+2(-1)^{r}S_{2}(r-1)\Big)\frac{1}{\omega }
\bigg],
\label{NLOADpole}
\end{eqnarray}
which is transformed into the following dispersion representation
\begin{eqnarray}
\hat{\gamma}^{\mathrm{LO}}(\M) &=&
4\bigg[-\sum_{j=1}^{\infty}\frac{1}{j+M}
+S_1^\infty
\bigg],\qquad\qquad S_1^\infty\equiv S_1(\infty)=\sum_{j=1}^{\infty}\frac{1}{j}\,,
\label{LOADpolep}\\
\hat{\gamma}^{\mathrm{NLO}}(M) &=&
8\sum_{j=1}^{\infty}
\bigg[
\frac{1+(-1)^{j}}{(j+\M)^{3}}
-2\,\frac{1+(-1)^{j}}{(j+\M)^{2}} S_{1}(j-1) \nonumber \\
&&\hspace{15mm} -\frac{(1+(-1)^{j})\zeta (2)+2(-1)^{j}S_{2}(j-1)}{j+M}
+\frac{3}{2}\z3+\z2 S_1^{\infty}\bigg],
\label{NLOADdisp}
\end{eqnarray}
where the last two terms are fixed at $\M=0$.

The disadvantage of this representation is its numerical values for positive integers~$\M$, where the harmonic sums are rational numbers. However, it provides the simplest way to study the analytic properties of anomalous dimension, for example, to study the relation between DGLAP and BFKL equations in $\cN=4$ SYM theory~\cite{Kotikov:2002ab}. 
Having the database for the discussed representation for the relevant harmonic sums, which is given in a wide range in Ref.~\cite{Velizhanin:2020avm}, we can study the properties of the anomalous dimension, related with the BFKL and the double-logarithmic equations.
Moreover, as we have such representation we can write the expression for the anomalous dimension of twist-2 operators either through the harmonic sums with the integer positive argument $\M$ either through dispersion representation, which provide information about  residues of these analytically continued functions. For the last case we will work with the results from the BFKL and the double-logarithmic equations. 
The solutions of these equations give the \emph{exact results} for the certain terms in expansion for the anomalous dimension over~$\omega$ (similar to Eqs.~(\ref{LOADpolep}) or (\ref{NLOADdisp})) in \emph{all orders} of perturbative theory.

\section{Double-logarithmic equation}\label{Section:2}

There are several ways to obtain the double-logarithmic asymptotic for the scattering amplitudes and related quantities.
For the first time it was obtained in QED in Refs.~\citep{Gorshkov:1966qd,Gorshkov:1966ht}, following Sudakov~\cite{Sudakov:1954sw}, which we briefly discuss below.

For an arbitrary two particle process with the momentum of the initial and final particles as $p_1$, $p_2$ and $p_1'$, $p_2'$ correspondingly, the usual invariants can be written as:
\begin{equation}
s=(p_1+p_2)^2,\qquad
t=q^2=(p_1-p_1')^2,\qquad
u=(p_1-p_2')^2.
\end{equation}
Following Sudakov~\cite{Sudakov:1954sw}, the asymptotics of the graphs, when $s\to\infty$ and $t$ or $u\sim 1$ can be computed with the help of expansion of all intermediate integrated four-momentum $k$ over transverse and longitudinal parts:
\begin{equation}
k=p_2\alpha+p_1\beta+k_\bot.
\end{equation}
$s$-dependence of the graph is contained in the integration over $\alpha$ and $\beta$. One integration is spent to the calculation of the residues in accordance with path poles for Feynamn graphs, while the second integration gives at most one $\ln s$. The remaining integral over $d^2k_\bot$ corresponds to a two-dimensional integral and for the scalar particles with the finite mass this integral is always finite and is some definite function of $t$.

However, in the presence of numerator or if the particle is massless the integral over $d^2k_\bot$ can diverge logarithmically for the large over small $k_\bot$ correspondingly. If the initial integral does not contain any divergences, this means that the integration over $k_\bot$ is cut-off by $s$ for large $k_\bot$ and $1/s$ for small $k_\bot$. The integration over the angle can be performed immediately, the remaining integration over $d^2k_\bot$ gives the second $\ln s$, that is, the double-logarithmic term appears.

As was shown in Ref.~\cite{Gorshkov:1966ht} the ladder graph of $2n+2$ order gives the double-logarithmic contribution, which is given by the following equation:
\begin{eqnarray}
F_n&=&f_0\left(\frac{\alpha}{2\pi}\right)^n\!\!
\int\limits^1_{1/s}\frac{d\alpha_n}{\alpha_n}\!
\int\limits^{\alpha_n}_{1/s}\frac{d\alpha_{n-1}}{\alpha_{n-1}}\ldots\!
\int\limits^{\alpha_2}_{1/s}\frac{d\alpha_1}{\alpha_1}\!
\int\limits^1_{1/s}\frac{d\beta_n}{\beta_n}\!
\int\limits^1_{\beta_{n}}\frac{d\beta_{n-1}}{\beta_{n-1}}\ldots\!
\int\limits^1_{\beta_2}\frac{d\beta_1}{\beta_1}\equiv f_0 J_N\,, \label{LadLn}\\
&&\hspace*{45mm} {s\alpha_i\beta_i\gg 1}\,,\nonumber
\end{eqnarray}
where $f_0$ is the result for the corresponding tree amplitude. $J_N$ can be found as the solution of the Bethe-Salpeter equation and the following expression for the double-logarithmic asymptotic of amplitude can be written:
\begin{equation}\label{DLBesselAs}
F(\ln s)=f_0\frac{2}{x}I_1(x),\qquad
x^2=\frac{2\alpha}{\pi}\ln^2\!s
\end{equation}
where  $I_1(x)$ is the modified Bessel function of the first kind.
The small-$x$ expansion of this solution looks like
\begin{equation}
F(\rho)=2 f_0\bigg(
\frac{1}{2}
+\frac{x^2}{16}
+\frac{x^4}{384}
+\frac{x^6}{18432}
+\frac{x^8}{1474560}
+\frac{x^{10}}{176947200}
\bigg)
.\label{SmallxBessel}
\end{equation}

\begin{figure}[t]
\begin{center}
\includegraphics[trim = 0mm 0mm 0mm 0mm, clip,scale=.32]{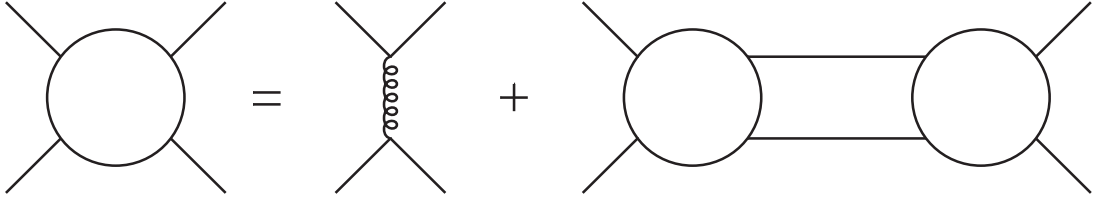}
\end{center}
\caption{
Equation for the double-logarithmic amplitudes.}\label{KL_Fig5_131}
\end{figure}

Another, more general approach, based on the infra-red evaluation equation, allows one to obtain the functional equation for the amplitude, or, to be more precise, for the partial wave in the space of a complex angular momentum $j$ not only in QED, but also in QCD~\cite{Kirschner:1982xw,Kirschner:1982qf,Kirschner:1983di}.
Graphically, this equation can be represented in the form given in Fig.~\ref{KL_Fig5_131}
for the simplest and the most important case for us.
The blob denotes the amplitude, which is infinity ladder with gluons rungs in the case of study the double-logarithmic asymptotic in QCD.
The main idea of this approach is the isolation of the softest intermediate momentum of quark lines on the rightmost diagram. Performing the Sudakov decomposition for this softest momentum and substituting the amplitudes by corresponding Sommerfeld-Watson integral it is possible to perform integration over Sudakov variables~\cite{Kirschner:1983di}. The final equation for the colour singlet amplitude with positive signature has the following form~\cite{Kirschner:1983di}:
\begin{equation}
f^+_0(\omega)=\frac{a_0g^2}{\omega}+\frac{1}{8\pi^2}\frac{1}{\omega}\left(f^+_0(\omega)\right)^2\label{4.1}
\end{equation}
with the boundary conditions
\begin{equation}
f^+_i(\omega)\Big|_{\omega\to 0}=\frac{a_0g^2}{\omega}\,,\qquad\qquad a_0=\frac{N_c^2-1}{2N_c}\,.
\end{equation}
The amplitudes with colour singlet exchange are given by
\begin{equation}\label{4.5}
f^+_0(\omega) = 4\pi^2\omega\left(1-\sqrt{1 -\frac{g^2(N_c^2-1)}{4\pi^2 N_c\omega^2}}\right)
\end{equation}
and $f^+_0$ has a square root cut starting from
\begin{equation}\label{4.6}
\omega=\omega^+_0\equiv\left(\frac{g^2( N_c^2-1)}{4\pi^2 N_c}\right)^{1/2}\,.
\end{equation}

This result was used in Refs.~\cite{Ermolaev:1995fx,Blumlein:1995jp} to find a similar equation for the non-singlet structure function at small $x$. For the non-singlet anomalous dimension of twist-2 operators in QCD the double-logarithmic equation is usually written as~\cite{Kirschner:1983di}
\begin{equation}
\gamma\big(\gamma+\omega\big)=-2\,C_F a_s\,,\qquad C_F=\frac{N_c^2-1}{2N_c}\,,\qquad 
a_s=\frac{\alpha_s}{4\pi}=\frac{g^2}{16\pi^2}
\end{equation}
for the expansion of the anomalous dimension at $j=0+\omega$.

\section{Generalised double-logarithmic equation}\label{Section:3}

The study of the analytical properties of the anomalous dimension of \mbox{twist-2} operators in $\cN=4$ SYM theory led to the suggestion about a simple generalisation of the original double-logarithmic equation~\cite{Kirschner:1982qf,Kirschner:1982xw,Kirschner:1983di}
\begin{eqnarray}\label{DL}
&&\ \gamma\,(2\,\omega+\gamma)=-16\,g^2\,.
\end{eqnarray}
The main idea was that in Eq.~(\ref{DL}) the corrections to the leading order equation will modify only the right-hand side and that such modification admit, besides the expansion over the coupling constant $g^2$, only the appearance of the regular terms in $\omega$ and, possibly, in $\gamma$. Such work was started by L.~Lipatov and A.~Onishchenko in 2004 for the general even $\M=-2,\ -4,\ -6,\ldots$, but was not published, then, some improvements of the double-logarithmic equation~(\ref{DL}) were proposed by L.~Lipatov in~\cite{Kotikov:2007cy}. However, it was not clear how to extend this procedure to higher orders. A simple generalisation of the original double-logarithmic equation~(\ref{DL}) was found in Ref.~\cite{Velizhanin:2011pb} for $\M=-2$ case. Substituting the results for the analytic continuation of the anomalous dimension of twist-2 operators (which was known at that moment up to five loops in planar $\cN=4$ SYM theory~\cite{Kotikov:2002ab,Kotikov:2003fb,Kotikov:2004er,Staudacher:2004tk,Kotikov:2007cy,Bajnok:2008qj,Lukowski:2009ce}) near $\M=-2+\omega$ into Eq.~(\ref{DL}), we obtained the following form of the generalised double-logarithmic equation:
\begin{eqnarray}\label{DLgenerC}
&&\ \gamma\,(2\,\omega+\gamma)=\sum_{k=1}\sum_{m=0}{\mathfrak C}_m^k\,\omega^m\,g^{2k}\,.
\end{eqnarray}
It turned out that the right-hand side of the modified double-logarithmic equation does not contain poles in~$\omega$.
The coefficients ${\mathfrak C}_m^k$ can be read directly from the Appendix~\ref{Section:DLgenerALL}  up to $g^{14}$ and up to transcendental numbers $\zeta_i$ with weight $12$. Here we write down the first four orders up to weight $6$\footnote{The math-file with full result can be found in the ancillary files of the arXiv version this paper}:
\begin{eqnarray}\label{DLgener}
\gamma\,(2\,\omega+\gamma)&=&16g^2 \Big[-1+\omega+(1+\z2) \omega ^2+(1-\z3) \omega ^3\nonumber\\&&\qquad\qquad \qquad \qquad \qquad
+(1+\z4) \omega ^4+(1-\z5) \omega ^5+
(1+\z6) \omega^6
\Big]\nonumber\\[2mm]&&\quad
+\,g^4 \bigg[
-64\, \z2
+\omega  (128\, \z2+96\, \z3)
+\,\omega ^2 (192\, \z2-160\, \z3-8 \,\z4)\nonumber\\&&\qquad\
+\,\omega ^3 (-256\, \z2 \z3+256\, \z2-224\, \z3+152\, \z4+360\, \z5)\nonumber\\&&\qquad\
+\,\omega ^4 \left(320\, \z2+144\, \zeta_3^2-288\, \z3+216\, \z4-144\, \z5+\frac{58}{3}\z6\right)
\bigg]\nonumber\\[2mm]&&\quad
+\,g^6 \bigg[
128\, \z3+256\, \z4
+\omega  (1152\, \z2 \z3+512\, \z3+672\, \z4-960\, \z5)\nonumber\\&&\qquad\
+\,\omega ^2 \left(384\, \z3-2688\, \z2 \z3-1056\, \zeta_3^2+256\, \z4+1504\, \z5-\frac{5000}{3}\z6\right)
\bigg]\nonumber\\[2mm]&&\quad
+\,g^8 \bigg[
2560\, \z2 \z3+384\, \zeta_3^2-128\, \z5+\frac{1888}{3}\z6
\bigg]+\ldots .\qquad\quad
\end{eqnarray}

In general, we can expect that the right-hand side of this equation~(\ref{DLgenerC}) should contain all special numbers from 1 to the most possible transcendental number ($\zeta_i$ or its generalisation).
Surprisingly, some of such terms are absent in the higher orders of the perturbative expansion, moreover, there is some regularity in such behaviour.
To make this statement clear, we have arranged the result for the right-hand side of the equation~(\ref{DLgenerC}) in the form of Table~\ref{Table:DLgener}, where all columns except the last correspond to the order of perturbative theory, while each row corresponds to the expected structure.
The last column presents the structure of the double-logarithmic term in the expression for analytically continued anomalous dimension at $\M=-2+\omega$, which is generated by the term from the corresponding row on the right-hand side of Eq.~(\ref{DLgenerC}) after it has been solved.
The vector $\zzeta_i$ means, that we include all $\zeta_{i_1,i_2,i_3,\ldots}$ and $\zeta_{i_1}\times\zeta_{i_2}\times\zeta_{i_3}\cdots$ inside such vector $\zzeta_i$ with given total transcendentality $i$, which is equal to the sums of absolute values of indices $|i_{1}|+|i_{2}|+|i_{3}|+\ldots$; for example, for transcendentality $i=7$ we have
\begin{equation}\label{zetaweight}
\zzeta_7=\left\{\z7,\z5\z2,\z3\z4
\right\}\,.
\end{equation}
The transcendentality of $\omega$ is equal to $(-1)$ as the transcendentality of the most simple harmonic sum $S_1$ is equal to $1$ and
\begin{equation}\label{TranscOmega}
S_1(-2+\omega)=\frac{1}{\omega}+{\mathcal{O}}(1)\,.
\end{equation}
Each cell in the table should respect the maximal transcendentality principle~\cite{Kotikov:2002ab}, that is, the total transcendentality of the corresponding expression should not exceed $(2\ell-1 +1)$ for the $\ell$ order of perturbative theory, where the additional $(+1)$ comes from the left-hand side $\gamma\,\omega$  in equation~(\ref{DLgenerC}).

\begin{table}
\tiny
\begin{equation*}
\begin{array}{cccccccccc}
\qquad  g^2 \qquad  &\qquad g^4 \qquad &\qquad g^6 \qquad &\qquad g^8 \qquad &\qquad g^{10} \qquad & \qquad g^{12} \qquad  &  &  &\quad  &  \displaystyle\left(\frac{g^2}{\omega^2}\right)^k\times \\[2mm]
{1}     &    0    &    0    &    0    &    0                                   & 0 &  &  &  &  \omega  \\[.1mm]
{\omega}  &    0    &    0    &    0    &    0                                   & 0 &  &  &  &  \omega^2    \\[.1mm]
{\omega^2}  & \boxed{1}  &    0    &    0    &    0                                      & 0 &  &  &  &  \omega^3  \\[.1mm]
{\zzeta_2\,\omega^2}  & {\zeta_2}  &    0    &    0    &    0                       & 0 &  &  &  &  \zzeta_2\,\omega^3  \\[.1mm]
{\omega^3}  & \boxed{\omega}  &    0    &    0    &    0                                 & 0 &  &  &  &  \omega^4  \\[.1mm]
\zzeta_2\,\omega^3  & {\zzeta_2\,\omega}  &    0    &    0    &    0               & 0 &  &  &  &  \zzeta_2\,\omega^4  \\[.1mm]
{\zzeta_3\,\omega^3}  & {\zzeta_3\,\omega}  &    0    &    0    &    0               & 0 &  &  &  &  \zzeta_3\,\omega^4  \\[.1mm]
{\omega^4}  & \boxed{\omega^2}  & \boxed{1}  &    0    &    0                                    & 0 &  &  &  &  \omega^5   \\[.1mm]
\zzeta_2\,\omega^4  & {\zzeta_2\,\omega^2}  & \boxed{\zzeta_2}  &    0    &    0            & 0 &  &  &  &  \zzeta_2\,\omega^5  \\[.1mm]
\zzeta_3\,\omega^4  & {\zzeta_3\,\omega^2}  & {\zzeta_3}  &    0    &    0            & 0 &  &  &  &  \zzeta_3\,\omega^5  \\[.1mm]
{\zzeta_4\,\omega^4}  & {\zzeta_4\,\omega^2}  & {\zzeta_4}  &    0    &    0            & 0 &  &  &  &  \zzeta_4\,\omega^5  \\[.1mm]
{\omega^5}  & \boxed{\omega^3}  & \boxed{\omega}  &    0    &    0                               & 0 &  &  &  &  \omega^6   \\[.1mm]
\zzeta_2\,\omega^5  & {\zzeta_2\,\omega^3}  & \boxed{\zzeta_2\,\omega}  &    0    &    0    & 0 &  &  &  &  \zzeta_2\,\omega^6  \\[.1mm]
\zzeta_3\,\omega^5  & {\zzeta_3\,\omega^3}  & {\zzeta_3\,\omega}  &    0    &    0    & 0 &  &  &  &  \zzeta_3\,\omega^6  \\[.1mm]
\zzeta_4\,\omega^5  & {\zzeta_4\,\omega^3}  & \zzeta_4\,\omega  &    0    &    0    & 0 &  &  &  &  \zzeta_4\,\omega^6  \\[.1mm]
\zzeta_5\,\omega^5  & {\zzeta_5\,\omega^3}  & \zzeta_5\,\omega  &    0    &    0    & 0 &  &  &  &  \zzeta_5\,\omega^6  \\[.1mm]
\omega^6  & \boxed{\omega^4}  & \boxed{\omega^2}  &    \boxed{1}    &    0                               & 0 &  &  &  &  \omega^7   \\[.1mm]
\zzeta_2\,\omega^6  & \zzeta_2\,\omega^4  & \boxed{\zzeta_2\,\omega^2}  &  \boxed{\zzeta_2}  &    0    & 0 &  &  &  &  \zzeta_2\,\omega^7  \\[.1mm]
\zzeta_3\,\omega^6  & \zzeta_3\,\omega^4  & \zzeta_3\,\omega^2  &  \boxed{\zzeta_3}  &    0    & 0 &  &  &  &  \zzeta_3\,\omega^7  \\[.1mm]
\zzeta_4\,\omega^6  & \zzeta_4\,\omega^4  & \zzeta_4\,\omega^2  &  \boxed{\zzeta_4}  &    0    & 0 &  &  &  &  \zzeta_4\,\omega^7  \\[.1mm]
\zzeta_5\,\omega^6  & \zzeta_5\,\omega^4  & \zzeta_5\,\omega^2  &  \zzeta_5  &    0    & 0 &  &  &  &  \zzeta_5\,\omega^7  \\[.1mm]
\zzeta_6\,\omega^6  & \zzeta_6\,\omega^4  & \zzeta_6\,\omega^2  &  \zzeta_6  &    0    & 0 &  &  &  &  \zzeta_6\,\omega^7  \\[.1mm]
\omega^7  & \boxed{\omega^5}  & \boxed{\omega^3}  &    \boxed{\omega}    &    0                               & 0 &  &  &  &  \omega^8   \\[.1mm]
\zzeta_2\,\omega^7  & \zzeta_2\,\omega^5  & \boxed{\zzeta_2\,\omega^3}  &  \boxed{\zzeta_2\,\omega}  &    0    & 0 &  &  &  &  \zzeta_2\,\omega^8  \\[.1mm]
\zzeta_3\,\omega^7  & \zzeta_3\,\omega^5  & \zzeta_3\,\omega^3  &  \boxed{\zzeta_3\,\omega}  &    0    & 0 &  &  &  &  \zzeta_3\,\omega^8  \\[.1mm]
\zzeta_4\,\omega^7  & \zzeta_4\,\omega^5  & \zzeta_4\,\omega^3  &  \boxed{\zzeta_4\,\omega}  &    0    & 0 &  &  &  &  \zzeta_4\,\omega^8  \\[.1mm]
\zzeta_5\,\omega^7  & \zzeta_5\,\omega^5  & \zzeta_5\,\omega^3  &  \zzeta_5\,\omega  &    0    & 0 &  &  &  &  \zzeta_5\,\omega^8  \\[.1mm]
\zzeta_6\,\omega^7  & \zzeta_6\,\omega^5  & \zzeta_6\,\omega^3  &  \zzeta_6\,\omega  &    0    & 0 &  &  &  &  \zzeta_6\,\omega^8  \\[.1mm]
\zzeta_7\,\omega^7  & \zzeta_7\,\omega^5  & \zzeta_7\,\omega^3  &  \zzeta_7\,\omega  &    0    & 0 &  &  &  &  \zzeta_7\,\omega^8  \\[.1mm]
\omega^8  & \boxed{\omega^6}  & \boxed{\omega^4}  &    \boxed{\omega^2}    &    \boxed{1}                               & 0 &  &  &  &  \omega^9   \\[.1mm]
\zzeta_2\,\omega^8  & \zzeta_2\,\omega^6  & \boxed{\zzeta_2\,\omega^4}  &  \boxed{\zzeta_2\,\omega^2}  &    \boxed{\zzeta_2}    & 0 &  &  &  &  \zzeta_2\,\omega^9  \\[.1mm]
\zzeta_3\,\omega^8  & \zzeta_3\,\omega^6  & \zzeta_3\,\omega^4  &  \boxed{\zzeta_3\,\omega^2}  &    \boxed{\zzeta_3}    & 0 &  &  &  &  \zzeta_3\,\omega^9 \\[.1mm]
\zzeta_4\,\omega^8  & \zzeta_4\,\omega^6  & \zzeta_4\,\omega^4  &  \boxed{\zzeta_4\,\omega^2}  &    \boxed{\zzeta_4}    & 0 &  &  &  &  \zzeta_4\,\omega^9  \\[.1mm]
\zzeta_5\,\omega^8  & \zzeta_5\,\omega^6  & \zzeta_5\,\omega^4  &  \zzeta_5\,\omega^2  &    \boxed{\zzeta_5}    & 0 &  &  &  &  \zzeta_5\,\omega^9  \\[.1mm]
\zzeta_6\,\omega^8  & \zzeta_6\,\omega^6  & \zzeta_6\,\omega^4  &  \zzeta_6\,\omega^2  &    \boxed{\zzeta_6}    & 0 &  &  &  &  \zzeta_6\,\omega^9  \\[.1mm]
\zzeta_7\,\omega^8  & \zzeta_7\,\omega^6  & \zzeta_7\,\omega^4  &  \zzeta_7\,\omega^2  &    {\zzeta_7}    & 0 &  &  &  &  \zzeta_7\,\omega^9  \\[.1mm]
\zzeta_8\,\omega^8  & \zzeta_8\,\omega^6  & \zzeta_8\,\omega^4  &  \zzeta_8\,\omega^2  &    {\zzeta_8}    & 0 &  &  &  &  \zzeta_8\,\omega^9  \\[.1mm]
\omega^9  & \boxed{\omega^7}  & \boxed{\omega^5}  &    \boxed{\omega^3}    &    \boxed{\omega}                               & 0 &  &  &  &  \omega^{10}   \\[.1mm]
\zzeta_2\,\omega^9  & \zzeta_2\,\omega^7  & \boxed{\zzeta_2\,\omega^5}  &  \boxed{\zzeta_2\,\omega^3}  &    \boxed{\zzeta_2\omega}    & 0 &  &  &  &  \zzeta_2\,\omega^{10}  \\[.1mm]
\zzeta_3\,\omega^9  & \zzeta_3\,\omega^7  & \zzeta_3\,\omega^5  &  \boxed{\zzeta_3\,\omega^3}  &    \boxed{\zzeta_3\omega}    & 0 &  &  &  &  \zzeta_3\,\omega^{10} \\[.1mm]
\zzeta_4\,\omega^9  & \zzeta_4\,\omega^7  & \zzeta_4\,\omega^5  &  \boxed{\zzeta_4\,\omega^3}  &    \boxed{\zzeta_4\omega}    & 0 &  &  &  &  \zzeta_4\,\omega^{10}  \\[.1mm]
\zzeta_5\,\omega^9  & \zzeta_5\,\omega^7  & \zzeta_5\,\omega^5  &  \zzeta_5\,\omega^3  &    \boxed{\zzeta_5\omega}    & 0 &  &  &  &  \zzeta_5\,\omega^{10}  \\[.1mm]
\zzeta_6\,\omega^9  & \zzeta_6\,\omega^7  & \zzeta_6\,\omega^5  &  \zzeta_6\,\omega^3  &    \boxed{\zzeta_6\omega}    & 0 &  &  &  &  \zzeta_6\,\omega^{10}  \\[.1mm]
\zeta_3^2\,\omega^9  & \zeta_3^2\,\omega^7  & \zeta_3^2\,\omega^5  &  \zeta_3^2\,\omega^3  &    \zeta_3^2\omega    & 0 &  &  &  &  \zeta_3^2\,\omega^{10}  \\[.1mm]
\zzeta_7\,\omega^9  & \zzeta_7\,\omega^7  & \zzeta_7\,\omega^5  &  \zzeta_7\,\omega^3  &    {\zzeta_7}\omega    & 0 &  &  &  &  \zzeta_7\,\omega^{10}  \\[.1mm]
\zzeta_8\,\omega^9  & \zzeta_8\,\omega^7  & \zzeta_8\,\omega^5  &  \zzeta_8\,\omega^3  &    {\zzeta_8}\omega    & 0 &  &  &  &  \zzeta_8\,\omega^{10}  \\[.1mm]
\zzeta_9\,\omega^9  & \zzeta_9\,\omega^7  & \zzeta_9\,\omega^5  &  \zzeta_9\,\omega^3  &    {\zzeta_9}\omega    & 0 &  &  &  &  \zzeta_9\,\omega^{10}  \\[.1mm]
\end{array}
\end{equation*}
\caption{The contributions to the generalised double-logarithmic equation~(\ref{DLgener}) and~(\ref{DLgenerALL}). The~terms marked by box drop out from the expected places in~(\ref{DLgener}) and~(\ref{DLgenerALL}).}.
\label{Table:DLgener}
\end{table}

\normalsize
From this table or directly from Eq.~(\ref{DLgenerC}) one can see, that some of the terms marked by box drop out from the expected places, which allow us to extend all-loop results, obtained in our previous paper~\cite{Velizhanin:2011pb}. Thus, all terms that do not depend on any special numbers $\zeta_i$ are contained in the first column, that is they appear only in the first order of perturbative expansion on the right-hand side of Eq.~(\ref{DLgenerC}). Therefore, only this term will generate all poles in the anomalous dimension of twist-2 operators, which do not contain any special numbers $\zeta_i$. For this case, we suppose that the equation~(\ref{DLgenerC}) has the following {\emph {exact}} form
\begin{equation}\label{DLgener1}
\gamma\,(2\,\omega+\gamma)\Big |_{\mathrm{Rational}}=16\,g^2\left(\sum_{n=1}^{\infty}\omega^n-1\right)=16\,g^2\left(\frac{1}{1-\omega}-2\right)\,.
\end{equation}
Surprisingly, the series in $\omega$ was resummed into a simple form.

The same is correct for other contributions. Thus, all poles of anomalous dimension with only the first transcendental number~$\zeta_2$ can be obtained from the solution of the following equation:
\begin{eqnarray}\label{DLgenerz2}
\gamma\,(2\,\omega+\gamma)\Big |_{\zeta_2}&=&16\,g^2\Bigg(\frac{1}{1-\omega}-2+\zeta_2\,\omega^2\Bigg)
-64\, g^4\, \zeta_2 \frac{\big(1-4\, \omega+2\, \omega^2\big)}{(1-\omega)^2}\quad\nonumber\\
&&\hspace*{-10mm}=16\,g^2\Bigg(\frac{1}{1-\omega}-2+\zeta_2\,\omega^2\Bigg) - 64\,g^4\,\zeta_2\left(2-\frac{1}{(1-\omega)^2}\right), \quad\end{eqnarray}
where the first two terms in the first bracket are the same as in Eq.~(\ref{DLgener1}), because we need to write it here (and in all examples below) to obtain the correct solution, which matches with the expression for the analytically continued seven-loop anomalous dimension.

Then, all poles of the anomalous dimension only with $\zeta_3$ can be obtained from the solution of the next equation:
\begin{eqnarray}\label{DLgenerz3}
\gamma\,(2\,\omega+\gamma)\Big |_{\zeta_3}
&=&16\,g^2\Bigg(\frac{1}{1-\omega}-2-\zeta_3\,\omega^3\Bigg)\nonumber\\&&
+32\,g^4\,\zeta_3
\Bigg(
\frac{3 - 11 \omega + 6 \omega ^2}{(1-\omega)^2}\,\omega
+4 g^2 \frac{1 + \omega - 6 \omega ^2}{(1-\omega)^3}
\Bigg)\nonumber\\
&&\hspace*{-15mm}=16\,g^2\Bigg(\frac{1}{1-\omega}-2-\zeta_3\,\omega^3\Bigg)
+32\,g^4\,\zeta_3\,\omega \left(6-\frac{1}{1-\omega}-\frac{2}{(1-\omega)^2}\right)\quad\nonumber\\&&
+128\,g^6\,\zeta_3 \left(-\frac{6}{1-\omega}+\frac{11}{(1-\omega)^2}-\frac{4}{(1-\omega)^3}\right)\,.
\end{eqnarray}

Using the seven-loop result for the planar anomalous dimension in $\cN=4$ SYM theory~\cite{Kotikov:2002ab,Kotikov:2003fb,Kotikov:2004er,Staudacher:2004tk,Kotikov:2007cy,Bajnok:2008qj,Lukowski:2009ce,Marboe:2014sya,Marboe:2016igj}, which is analytically continued at double-logs value up to weight~$12$~(\ref{DLgenerALL}), we have found, using \texttt{MATHEMATICA} function \texttt{FindGeneratingFunction}, the following resummed expression ($\barom=1-\omega$):
\allowdisplaybreaks
\begin{align}\label{DLgenerRessum}
\gamma\,(2\,\omega+\gamma)&=16\,g^2\left(\frac{1}{\barom}-2\right)
+\z2\,g^2\Bigg[16\,\omega^2 - 64\,g^2\left(2-\frac{1}{\barom^2}\right)\Bigg]\nonumber\\
&
+\z3\,g^2\Bigg[
-16\,\omega^3 
+32\,g^2\,\omega\left(6-\frac{1}{\barom}-\frac{2}{\barom^2}\right)
+128\,g^4\left(-\frac{6}{\barom}+\frac{11}{\barom^2}-\frac{4}{\barom^3}\right)
\Bigg]\nonumber\\
&
+\z4\,g^2\Bigg[
16\,\omega^4 
+8\,g^2\,\omega^2\left(-12+\frac{3}{\barom}+\frac{8}{\barom^2}\right)
+32\,g^4\left(-70+\frac{179}{\barom}-\frac{145}{\barom^2}+\frac{44}{\barom^3}\right)
\Bigg]\nonumber\\
&
+\z5\,g^2\Bigg[
-16\,\omega^5 
+16\,g^2\,\omega^3\left(\frac{55}{2}-\frac{1}{\barom}-\frac{4}{\barom^2}\right)
+32\,g^4\,\omega\left(40-\frac{219}{\barom}+\frac{181}{\barom^2}-\frac{32}{\barom^3}\right)
\nonumber\\&\qquad\qquad 
+128\,g^6\left(\frac{200}{\barom}-\frac{334}{\barom^2}+\frac{149}{\barom^3}-\frac{16}{\barom^4}\right)
\Bigg]\nonumber\\
&
+\z2\z3\,g^2\Bigg[
0\times\omega^5 
-256\,g^2\,\omega^3
+128\,g^4\,\omega\left(24-\frac{13}{\barom}+\frac{2}{\barom^2}-\frac{4}{\barom^3}\right)
\nonumber\\&\qquad\qquad 
+512\,g^6\left(\frac{6}{\barom}-\frac{18}{\barom^2}+\frac{29}{\barom^3}-\frac{12}{\barom^4}\right)
\Bigg]\nonumber\\
&
+\z6\,g^2\Bigg[
16\,\omega^6
+2\,g^2\omega^4\left(-\frac{82}{3}+\frac{5}{\barom}+\frac{32}{\barom^2}\right)
+\frac{8}{3}\,g^4\omega^2\left(-3730+\frac{5564}{\barom}-\frac{3275}{\barom^2}+\frac{816}{\barom^3}\right)
\nonumber\\&\qquad\qquad 
+\frac{32}{3}\,g^6\left(11758-\frac{29372}{\barom}+\frac{28266}{\barom^2}-\frac{13005}{\barom^3}+\frac{2412}{\barom^4}\right)
\Bigg]\nonumber\\
&
+\zeta_3^2\,g^2\Bigg[
0\times\omega^6
+144\,g^2\omega^4
+32\,g^4\omega^2\left(-66+\frac{19}{\barom}+\frac{6}{\barom^2}+\frac{8}{\barom^3}\right)
\nonumber\\&\qquad\qquad 
+128\,g^6\left(-32-\frac{16}{\barom}+\frac{448}{\barom^2}-\frac{580}{\barom^3}+\frac{192}{\barom^4}\right)
\nonumber\\&\qquad\qquad 
+1024\,g^8\left(-\frac{78}{\barom^2}+\frac{181}{\barom^3}-\frac{135}{\barom^4}+\frac{32}{\barom^5}\right)
\Bigg]\nonumber\\
&
+\z7\,g^2\Bigg[
-16\,\omega^7
+g^2\,\omega^5\left(777-\frac{6}{\barom}-\frac{64}{\barom^2}\right)
+4\,g^4\,\omega^3\left(-210-\frac{3792}{\barom}+\frac{2783}{\barom^2}-\frac{384}{\barom^3}\right)
\nonumber\\&\qquad\qquad 
+16\,g^6\,\omega\left(-10556+\frac{27728}{\barom}-\frac{22661}{\barom^2}+\frac{6618}{\barom^3}-\frac{576}{\barom^4}\right)\nonumber\\&\qquad\qquad
+128\,g^8\left(-\frac{4424}{\barom}+\frac{7862}{\barom^2}-\frac{4359}{\barom^3}+\frac{895}{\barom^4}-\frac{64}{\barom^5}\right)
\Bigg]\nonumber\\
&
+\z2\z5\,g^2\Bigg[
0\times\omega^7
-384\,g^2\,\omega^5
+32\,g^4\,\omega^3\left(345-\frac{126}{\barom}+\frac{12}{\barom^2}-\frac{16}{\barom^3}\right)
\nonumber\\&\qquad\qquad 
+256\,g^6\,\omega\left(-364+\frac{482}{\barom}-\frac{393}{\barom^2}+\frac{242}{\barom^3}-\frac{48}{\barom^4}\right)\nonumber\\&\qquad\qquad
+1024\,g^8\left(-\frac{30}{\barom}+\frac{222}{\barom^2}-\frac{436}{\barom^3}+\frac{239}{\barom^4}-\frac{32}{\barom^5}\right)
\Bigg]\nonumber\\
&
+\z4\z3\,g^2\Bigg[
0\times\omega^7
-280\,g^2\,\omega^5
+16\,g^4\,\omega^3\left(-50+\frac{53}{\barom}-\frac{38}{\barom^2}-\frac{32}{\barom^3}\right)
\nonumber\\&\qquad\qquad 
+64\,g^6\,\omega\left(728-\frac{360}{\barom}-\frac{936}{\barom^2}+\frac{1041}{\barom^3}-\frac{360}{\barom^4}\right)\nonumber\\&\qquad\qquad
+512\,g^8\left(-\frac{792}{\barom}+\frac{2665}{\barom^2}-\frac{3272}{\barom^3}+\frac{1869}{\barom^4}-\frac{432}{\barom^5}\right)
\Bigg]\nonumber\\
&
+\z8\,g^2\Bigg[
16\,\omega^8
+\frac{1}{2}g^2\,\omega^6\left(-\frac{842}{15}+\frac{7}{\barom}+\frac{128}{\barom^2}\right)
\nonumber\\&\qquad\qquad 
+\frac{2}{45}\,g^4\,\omega^4\left(-444272+\frac{643676}{\barom}-\frac{334419}{\barom^2}+\frac{66240}{\barom^3}\right)
\nonumber\\&\qquad\qquad 
+\frac{8}{45}\,g^6\,\omega^2\left(-2893674+\frac{6248681}{\barom}-\frac{5371295}{\barom^2}+\frac{2145970}{\barom^3}-\frac{335520}{\barom^4}\right)\nonumber\\&\qquad\qquad
+g^8\Big(\cdots\Big)
\Bigg]\nonumber\\
&
+\zeta_{53}\,g^2\Bigg[
0\times\omega^8
-\frac{144}{5}g^2\,\omega^6
+\frac{192}{5}\,g^4\,\omega^4\left(34-\frac{27}{\barom}+\frac{3}{\barom^2}\right)
\nonumber\\&\qquad\qquad 
+\frac{128}{5}\,g^6\,\omega^2\left(-698+\frac{1177}{\barom}-\frac{600}{\barom^2}+\frac{90}{\barom^3}\right)\nonumber\\&\qquad\qquad
+\frac{256}{45}\,g^8\left(-\frac{3498}{\barom}-\frac{10727}{\barom^2}-\frac{12606}{\barom^3}+\frac{6740}{\barom^4}-\frac{1384}{\barom^5}\right)
\Bigg]\nonumber\\
&
+\z5\z3\,g^2\Bigg[
0\times\omega^8
+272\,g^2\,\omega^6
+32\,g^4\,\omega^4\left(-189-\frac{43}{\barom}+\frac{26}{\barom^2}+\frac{16}{\barom^3}\right)
\nonumber\\&\qquad\qquad 
+64\,g^6\,\omega^2\left(124-\frac{999}{\barom}-\frac{260}{\barom^2}+\frac{1042}{\barom^3}-\frac{288}{\barom^4}\right)\nonumber\\&\qquad\qquad
+g^{8}\Big(\cdots\Big)+g^{10} \Big(\cdots\Big)
\Bigg]\nonumber\\
&
+\z2\zeta_{3}^2\,g^2\Bigg[
0\times\omega^8
+0\times g^2\,\omega^6
+96\,g^4\,\omega^4\left(45-\frac{8}{\barom}\right)
\nonumber\\&\qquad\qquad 
+256\,g^6\,\omega^2\left(228-\frac{127}{\barom}-\frac{16}{\barom^2}+\frac{6}{\barom^3}-\frac{12}{\barom^3}\right)\nonumber\\&\qquad\qquad
+g^{8}\Big(\cdots\Big)+g^{10} \Big(\cdots\Big)
\Bigg]\nonumber\\
&
+\z9\,g^2\Bigg[
-16\,\omega^9
+g^2\,\omega^7\left(-1251+\frac{2}{\barom}+\frac{64}{\barom^2}\right)
\nonumber\\&\qquad\qquad 
+\frac{4}{9}\,g^4\,\omega^5\left(25258+\frac{57433}{\barom}-\frac{40473}{\barom^2}+\frac{4608}{\barom^3}\right)
\nonumber\\&\qquad\qquad 
+g^6\omega^3\Big(\cdots\Big)+g^8\omega\Big(\cdots\Big)+g^{10}\Big(\cdots\Big)
\Bigg]\nonumber\\
&
+\zeta_3^3\,g^2\Bigg[
0\times\omega^9
+0\times g^2\,\omega^7
+32\,g^4\,\omega^5\left(-55+\frac{8}{\barom}\right)
\nonumber\\&\qquad\qquad 
+64\,g^6\,\omega^3\left(-450+\frac{263}{\barom}-\frac{76}{\barom^2}+\frac{80}{\barom^3}\right)\nonumber\\&\qquad\qquad
+g^8\,\omega\Big(\cdots\Big)+g^{10}\Big(\cdots\Big)+g^{12}\Big(\cdots\Big)
\Bigg]\nonumber\\
&
+\z2\z7\,g^2\Bigg[
0\times\omega^9
-512\times g^2\,\omega^7
+8\,g^4\,\omega^5\left(-3108+\frac{771}{\barom}-\frac{58}{\barom^2}+\frac{64}{\barom^3}\right)
\nonumber\\&\hspace*{20mm}
+g^6\,\omega^3\Big(\cdots\Big)
+g^8\,\omega\Big(\cdots\Big)
+g^{10}\Big(\cdots\Big)
\Bigg]\nonumber\\
&
+\z4\z5\,g^2\Bigg[
0\times\omega^9
-408\times g^2\,\omega^7
+8\,g^4\,\omega^5\left(-214+\frac{211}{\barom}+\frac{6}{\barom^2}+\frac{64}{\barom^3}\right)
\nonumber\\&\hspace*{20mm}
+g^6\,\omega^3\Big(\cdots\Big)
+g^8\,\omega\Big(\cdots\Big)
+g^{10}\Big(\cdots\Big)
\Bigg]\nonumber\\
&
+\z6\z3\,g^2\Bigg[
0\times\omega^9
-266\times g^2\,\omega^7
+\frac{4}{3}\,g^4\,\omega^5\left(9899-\frac{5905}{\barom}+\frac{798}{\barom^2}+\frac{384}{\barom^3}\right)
\nonumber\\&\hspace*{20mm}
+g^6\,\omega^3\Big(\cdots\Big)
+g^8\,\omega\Big(\cdots\Big)
+g^{10}\Big(\cdots\Big)
\Bigg]
\ \ +\ \ \cdots\,,
\end{align}
where all terms with transcedentality $10$ and more or denoting by $(\cdots)$ cannot be resummed with the available data.
The above equation again has some regularity, which can be seen from Table~\ref{Table:RGDL}. As the most interesting property, the maximal negative power of $\barom$ is equal to $\ell$ for $g^{2\ell}$ contribution. 
\begin{table}[h]
\begin{equation*}
\begin{array}{c|c|c|c|c|cc|cc|ccc|cccc|ccc}
    &              &\ \z2\ &\ \z3\ &\ \z4\ &\ \z5\ &\z2\z3\ &\ \z6\ &\zeta_3^2\ &\ \z7\ & \z2\z5\ & \z4\z3\ &\ \z8\ & \zeta_{5,3}\ & \z3\z5\ & \z2\zeta_3^2\ &\ \z9\ &\z2\z7\ & \zeta_3^3\ \\[1mm]
 \hline
\ g^2\ &  1           &\tx&\tx&\tx&\tx&   &\tx&   &\tx&   &   &\tx&   &   &   &\tx&   &    \\[.1mm]
 \hline
    &  1              &\tx&\tx&\tx&\tx&\tx&\tx&\tx&\tx&\tx&\tx&\tx&\tx&\tx&   &\tx&\tx&    \\[.1mm]
\ g^4\ &\ 1/{\barom}\ &   &\tx&\tx&\tx&   &\tx&   &\tx&   &   &\tx&   &   &   &\tx&   &    \\[.1mm]
   &\ 1/{\barom^{2}}\ &\tx&\tx&\tx&\tx&   &\tx&   &\tx&   &   &\tx&   &   &   &\tx&   &    \\[.1mm]
 \hline
    &  1              &   &   &\tx&\tx&\tx&\tx&\tx&\tx&\tx&\tx&\tx&\tx&\tx&\tx&\tx&\tx&\tx \\[.1mm]
    &1/{\barom}       &   &\tx&\tx&\tx&\tx&\tx&\tx&\tx&\tx&\tx&\tx&\tx&\tx&\tx&\tx&\tx&\tx \\[.1mm]
\ g^6\ &1/{\barom^{2}}&   &\tx&\tx&\tx&\tx&\tx&\tx&\tx&\tx&\tx&\tx&\tx&\tx&   &\tx&\tx&    \\[.1mm]
    &1/{\barom^{3}}   &   &\tx&\tx&\tx&\tx&\tx&\tx&\tx&\tx&\tx&\tx&   &\tx&   &\tx&\tx&    \\[.1mm]
 \hline
    &  1              &   &   &   &   &   &\tx&\tx&\tx&\tx&\tx&\tx&\tx&\tx&\tx&\tx&\tx&\tx \\[.1mm]
    &1/{\barom}       &   &   &   &\tx&\tx&\tx&\tx&\tx&\tx&\tx&\tx&\tx&\tx&\tx&\tx&\tx&\tx \\[.1mm]
\ g^8\ &1/{\barom^{2}}&   &   &   &\tx&\tx&\tx&\tx&\tx&\tx&\tx&\tx&\tx&\tx&\tx&\tx&\tx&\tx \\[.1mm]
    &1/{\barom^{3}}   &   &   &   &\tx&\tx&\tx&\tx&\tx&\tx&\tx&\tx&\tx&\tx&\tx&\tx&\tx&\tx \\[.1mm]
    &1/{\barom^{4}}   &   &   &   &\tx&\tx&\tx&\tx&\tx&\tx&\tx&\tx&   &\tx&\tx&\tx&\tx&\tx \\[.1mm]
 \hline
    &  1              &   &   &   &   &   &   &   &   &   &   &   &   &\tx&\tx&\tx&\tx&\tx \\[.1mm]
    &1/{\barom}       &   &   &   &   &   &   &   &\tx&\tx&\tx&\tx&\tx&\tx&\tx&\tx&\tx&\tx \\[.1mm]
\ g^{10}\ &1/{\barom^{2}}& &  &   &   &   &   &\tx&\tx&\tx&\tx&\tx&\tx&\tx&\tx&\tx&\tx&\tx \\[.1mm]
    &1/{\barom^{3}}   &   &   &   &   &   &   &\tx&\tx&\tx&\tx&\tx&\tx&\tx&\tx&\tx&\tx&\tx \\[.1mm]
    &1/{\barom^{4}}   &   &   &   &   &   &   &\tx&\tx&\tx&\tx&\tx&\tx&\tx&\tx&\tx&\tx&\tx \\[.1mm]
    &1/{\barom^{5}}   &   &   &   &   &   &   &\tx&\tx&\tx&\tx&\tx&\tx&\tx&\tx&\tx&\tx&\tx \\[.1mm]
 \hline
    &  1              &   &   &   &   &   &   &   &   &   &   &   &   &   &   &   &   &\tx \\[.1mm]
    &1/{\barom}       &   &   &   &   &   &   &   &   &   &   &   &   &\tx&\tx&\tx&\tx&\tx \\[.1mm]
\ g^{12}\ &1/{\barom^{2}}& &  &   &   &   &   &   &   &   &   &   &   &\tx&\tx&\tx&\tx&\tx \\[.1mm]
    &1/{\barom^{3}}   &   &   &   &   &   &   &   &   &   &   &   &   &\tx&\tx&\tx&\tx&\tx \\[.1mm]
    &1/{\barom^{4}}   &   &   &   &   &   &   &   &   &   &   &   &   &\tx&\tx&\tx&\tx&\tx \\[.1mm]
    &1/{\barom^{5}}   &   &   &   &   &   &   &   &   &   &   &   &   &\tx&\tx&\tx&\tx&\tx \\[.1mm]
    &1/{\barom^{6}}   &   &   &   &   &   &   &   &   &   &   &   &   &\tx&\tx&\tx&\tx&\tx \\[.1mm]
 \hline
    &  1              &   &   &   &   &   &   &   &   &   &   &   &   &   &   &   &   &    \\[.1mm]
    &1/{\barom}       &   &   &   &   &   &   &   &   &   &   &   &   &   &   &   &   &\tx \\[.1mm]
\ g^{14}\ &1/{\barom^{2}}& &  &   &   &   &   &   &   &   &   &   &   &   &   &   &   &\tx \\[.1mm]
    &1/{\barom^{3}}   &   &   &   &   &   &   &   &   &   &   &   &   &   &   &   &   &\tx \\[.1mm]
    &1/{\barom^{4}}   &   &   &   &   &   &   &   &   &   &   &   &   &   &   &   &   &\tx \\[.1mm]
    &1/{\barom^{5}}   &   &   &   &   &   &   &   &   &   &   &   &   &   &   &   &   &\tx \\[.1mm]
    &1/{\barom^{6}}   &   &   &   &   &   &   &   &   &   &   &   &   &   &   &   &   &\tx \\[.1mm]
    &1/{\barom^{7}}   &   &   &   &   &   &   &   &   &   &   &   &   &   &   &   &   &\tx \\[.1mm]
\end{array}
\end{equation*}
\caption{The contribution on the right-hand side of the resummed generalised double-logarithmic equation~(\ref{DLgenerRessum}). Nonempty cells, labeled with ``$\times$'', are presented in Eq.~(\ref{DLgenerRessum}). The cells of the last five colomns for contributions higher than $g^8$ are expected according to our assumption.}
\label{Table:RGDL}
\end{table}

In the general case, the result for the right-hand side of the resummed generalised double-logarithmic equation can be written as:
\begin{eqnarray}
\label{DLgenerRessumALL}
\gamma\,(2\,\omega+\gamma)&=&16\,g^2\Bigg(\frac{1}{1-\omega}-2\Bigg)\nonumber\\
& &\hspace*{-15mm}
+{{\bm\zeta}}_{{\bm \ell}}\Bigg[
\sum_{i=1}^{\lfloor\frac{{\bm\ell}-1}{2}\rfloor}{g^{2(i+1)}}\omega^{{\bm\ell}-2i}
\sum_{k=0}^{i+1}\frac{C^{i,k}_{{{\bm\zeta}}_{{\bm \ell}}}}{(1-\omega)^k}
+{g^{2{\lfloor\frac{{\bm\ell}+3}{2}\rfloor}}}
\sum_{k=0}^{{\lfloor\frac{{\bm\ell}+3}{2}\rfloor}}\frac{C^{k}_{{{\bm\zeta}}_{{\bm \ell}}}}{(1-\omega)^k}\Bigg]
\nonumber\\&&\hspace*{-15mm}
+\zeta_3{\hat{\bm\zeta}}_{{\bm \ell}}\Bigg[
\sum_{i=1}^{\lfloor\frac{{\bm\ell}-1+2}{2}\rfloor}{g^{2(i+1)}}\omega^{{\bm\ell}-2i}
\sum_{k=0}^{i+1}\frac{C^{i,k}_{\zeta_3{\hat{\bm\zeta}}_{{\bm \ell}}}}{(1-\omega)^k}
+{g^{2{\lfloor\frac{{\bm\ell}+3+2}{2}\rfloor}}}
\sum_{k=0}^{{\lfloor\frac{{\bm\ell}+3+2}{2}\rfloor}}\frac{C^{k}_{\zeta_3{\hat{\bm\zeta}}_{{\bm \ell}}}}{(1-\omega)^k}\Bigg]+\ \cdots,\qquad
\end{eqnarray}
where ${{\bm\zeta}}_{{\bm \ell}}$ is one of $\{\zeta_{i},\ \zeta_{i_1,i_2,\ldots},\ \zeta_{2i_1}\zeta_{2i_2+1},\ \ldots\}$ with total weight equal to $\ell$ (see Eq.~(\ref{zetaweight}) and text around that equation), while ${\hat{\bm\zeta}}_{{\bm \ell}}$ is one of $\{\zeta_{2i+1},\ \zeta_{i_1,i_2,\ldots},\ \zeta_{2i_1}\zeta_{2i_2+1},\ \ldots\}$ .

To find other interesting properties of the resummed generalised double-logarithmic equation~(\ref{DLgenerRessum}), we give its expansion in powers of $g^2$ and $\barom$, which looks like:
\begin{align}
\gamma\left(2\,\omega+\gamma\right)&
=16\,g^2\Bigg[ -2+\,\sum_{i=2}(-1)^i\zeta_i\,\omega^i
+\frac{1}{\barom}\Bigg]
\nonumber\\&\hspace*{-8mm}
+g^4\Bigg[
-128 {\z2}
+192 \omega {\z3}
-96 \omega^2 {\z4}
+\omega^3 \big(
440 {\z5}
-256 {\z2} {\z3}
\big)
+\omega^4 \left(
144 \zeta_3^2
-\frac{164}{3} {\z6}
\right)
\nonumber\\&
+\omega^5 \big(
777 {\z7}
-384 {\z2} {\z5}
-280 {\z3} {\z4}
\big)
+\omega^6 \left(
272 {\z3} {\z5}
-\frac{144 }{5}{\zeta_{5,3}}
-\frac{421 }{15}{\z8}
-\frac{141}{7} \zeta_{7,3}
\right)
\nonumber\\&
+\omega^7 \big(
-512 {\z2} {\z7}
-1251 {\z9}
\big)
-640 \omega^9 {\z2} {\z9}
-\frac{46}{3} \omega^{10} \zeta_{9,3}
\nonumber\\&
+\frac{1}{\barom}
-\Bigg(
-32 \omega {\z3}
+24 \omega^2 {\z4}
-16 \omega^3 {\z5}
+10 \omega^4 {\z6}
-6 \omega^5 {\z7}
+\frac{7}{2} \omega^6 {\z8}
+2 \omega^7 {\z9}
\Bigg)
\nonumber\\&
+\frac{64}{\barom^2} \Bigg(
+ {\z2}
- \omega {\z3}
+ \omega^2 {\z4}
- \omega^3 {\z5}
+ \omega^4 {\z6}
- \omega^5 {\z7}
+ \omega^6 {\z8}
+ \omega^7 {\z9}
\Bigg)
\Bigg]
\nonumber\\&\hspace*{-8mm}
+g^6 \Bigg[
-2240 {\z4}
+\omega \big(
3072 {\z2} {\z3}
+1280 {\z5}
\big)
+\omega^2 \left(
-2112 \zeta_3^2
-\frac{29840}{3}{\z6}
\right)
\nonumber\\&
+\omega^3 \big(
11040 {\z2} {\z5}-
800 {\z3} {\z4}
-840 {\z7}
\big)
\nonumber\\&
+\omega^4 \left(
270 {\z2} \zeta_3^2
+10976 {\z2} {\z3} {\z5}
-6048 {\z3} {\z5}
+\frac{6528   }{5}{\zeta_{5,3}}
-\frac{888544 }{45}{\z8}
\right)
\nonumber\\&
+\omega^5 \left(
256 {\z2} \zeta_3^3
-3840 {\z2} {\z3} {\z5}
+18648 {\z2} {\z7}
-1760  \zeta_3^3+
\frac{101032 }{9}{\z9}
\right)
\nonumber\\&
+\omega^6 \big(
16 {\z2} {\z3} {\z5}
-576 {\z2} \zeta_{5,3}
\big)
+\omega^7 \left(
-\frac{2304}{5} {\z2} \zeta_{5,3}
+\frac{576}{5} \zeta_{3,5,3}
-768 {\z2} {\z3} {\z5}
\right)
\nonumber\\&
+\omega^8 \left(
-\frac{2304}{5} {\z2} \zeta_{5,3}
-\frac{2304}{5} \zeta_{3,5,3}
+192 \zeta_{4,4,2,2}
-768 {\z2} {\z3} {\z5}
-128 \zeta_3^4
\right)
\nonumber\\&
+\frac{1}{\barom} \Bigg(
-768 {\z3}
+5728 {\z4}
+\omega \big(
-1664 {\z2} {\z3}
-7008 {\z5}
\big)
\nonumber\\&
+\omega^2 \left(
608 \zeta_3^2
+\frac{44512 }{3}{\z6}
\right)
+\omega^3 \big(
-4032 {\z2} {\z5}
+848 {\z3} {\z4}
-15168 {\z7}
\big)
\nonumber\\&
+\omega^4 \left(
\frac{1287352}{45} {\z8}
-48 {\z2} \zeta_3^2
-1376 {\z3} {\z5}
-\frac{5184 }{5}{\zeta_{5,3}}
\right)
+\omega^5 \left(
256 \zeta_3^3
+\frac{229732}{9}{\z9}
\right)
\Bigg)
\nonumber\\&
+\frac{1}{\barom^2}
\Bigg(
1408 {\z3}
-4640 {\z4}
+\omega \big(
256 {\z2} {\z3}
+5792 {\z5}
\big)
+\omega^2 \left(
192 \zeta_3^2
-\frac{26200 }{3}{\z6}
\right)
\nonumber\\&
+\omega^3 \big(
384 {\z2} {\z5}
-608 {\z3} {\z4}
+11132 {\z7}
\big)
\nonumber\\&
+\omega^4 \left(
832\, {\z3}\, {\z5}
+\frac{576}{5} {\zeta_{53}}
-\frac{222946}{15} {\z8}
\right)
-17988\, \omega^5 {\z9}
\Bigg)
\nonumber\\&
+\frac{1}{\barom^3}\Bigg(
-512 {\z3}
+1408 {\z4}
+\omega (
-512 {\z2} {\z3}
-1024 {\z5}
)
+\omega^2 (
256 \zeta_3^2
+2176 {\z6}
)
\nonumber\\&
+\omega^3 (
-512 {\z2} {\z5}
-512 {\z3} {\z4}
-1536 {\z7}
)
+\omega^4 (
512 {\z3} {\z5}
+2944 {\z8}
)
+2048 \omega^5 {\z9}
\Bigg)
\Bigg]
\nonumber\\&\hspace*{-8mm}
+g^8\Bigg[
\frac{376256 {\z6}}{3}
-1024 \zeta_3^2
+\omega (
-93184 {\z2} {\z5}
+46592 {\z3} {\z4}
-168896 {\z7}
-168896 {\z9}
)
\nonumber\\&
+\omega^2 (
58368 {\z2} \zeta_3^2
+7936 {\z3} {\z5}
-\frac{89344 }{5}{\zeta_{5,3}}
-\frac{7716464 }{15}{\z8}
)
\nonumber\\&
+\omega^3 (
-1651840 {\z2} {\z7}
-28800 \zeta_3^3
)
\nonumber\\&
+\frac{1}{\barom} \Bigg(
+3072 {\z2} {\z3}
-512 \zeta_3^2
+25600 {\z5}
-\frac{939904}{3}{\z6}
\nonumber\\&
+\omega \big(
123392 {\z2} {\z5}
-23040 {\z3} {\z4}
+443648 {\z7}
+443648 {\z9}
\big)
\nonumber\\&
+\omega^2 \left(
-32512 {\z2} \zeta_3^2
-63936 {\z3} {\z5}
+\frac{150656}{5} {\zeta_{5,3}}
+\frac{49989448}{45} {\z8}
\right)
+16832 \omega^3 \zeta_3^3
\Bigg)
\nonumber\\&
+\frac{1}{\barom^2} \Bigg(
-9216 {\z2} {\z3}
+14336 \zeta_3^2
-42752 {\z5}
+301504 {\z6}
\nonumber\\&
+\omega \big(
-100608 {\z2} {\z5}
-59904 {\z3} {\z4}
-362576 {\z7}
-362576 {\z9}
\big)
\nonumber\\&
+\omega^2 \left(
-4096 {\z2} \zeta_3^2
-16640 {\z3} {\z5}
-15360 {\zeta{5,3}}
-\frac{8594072}{9} {\z8}
\right)
-4864 \omega^3 \zeta_3^3
\Bigg)
\nonumber\\&
+\frac{1}{\barom^3} \Bigg(
+14848 {\z2} {\z3}
-18560 \zeta_3^2
+19072 {\z5}
-138720 {\z6}
\nonumber\\&
+\omega \big(
61952 {\z2} {\z5}
+66624 {\z3} {\z4}
+105888 {\z7}
+105888 {\z9}
\big)
\nonumber\\&
+\omega^2 \left(
-1536 {\z2} \zeta_3^2
+48256 {\z3} {\z5}
+2304 {\zeta_{5,3}}
+\frac{3433552}{9} {\z8}
\right)
+5120 \omega^3 \zeta_3^3
\Bigg)
\nonumber\\&
+\frac{1}{\barom^4} \Bigg(
-6144 {\z2} {\z3}
+6144 \zeta_3^2
-2048 {\z5}
+25728 {\z6}
\nonumber\\&
+\omega \big(
-12288 {\z2} {\z5}
-23040 {\z3} {\z4}
-9216 {\z7}
-9216 {\z9}
\big)
\nonumber\\&
+\omega^2 \big(
-3072\z2\zeta_3^2
-18432\z3\z5
-59648 {\z8}\big)
\Bigg)
\Bigg]
\nonumber\\&\hspace*{-8mm}
+g^{10}\Bigg[
\Bigg(
7168 {\z2} \zeta_3^2
+3072 {\z3} {\z5}
+256\,\omega^4 (
1420 {\z2} \zeta_3^2
+283 \zeta_3^3
+582 {\z3} {\z5}
)
\nonumber\\&
+256\,\omega\big(
1312 {\z2} {\z7}
-308 {\z2} \zeta_3^2
-39 \zeta_3^3
-42 {\z3} {\z5}
\big)
\nonumber\\&
+256\,\omega^2 (
218 \zeta_3^3
-432 {\z2} \zeta_3^2
-1119 {\z3} {\z5}
)
+256\,\omega^3 (
390 \zeta_3^3
-76 {\z2}\zeta_3^2
-1457 {\z3} {\z5}
)
\Bigg)
\nonumber\\&
+\frac{32}{15}\frac{1}{\barom} \Bigg(
-14400 {\z2} {\z5}
-190080 {\z3} {\z4}
-9328 \zeta_{5,3}
-265440 {\z7}
+3235149\z8
\Bigg)
\nonumber\\&
+\frac{64}{45}\frac{1}{\barom^2} \Bigg(
-56160 {\zeta_3^2}
+159840 {\z2} {\z5}
+959400 {\z3} {\z4}
\nonumber\\&\hspace{25mm}
-42908\zeta_{5,3}
+707580 {\z7}
-15485539\z8
\Bigg)
\nonumber\\&
+\frac{32}{45}\frac{1}{\barom^3} \Bigg(
260640 {\zeta_3^2}
-627840 {\z2} {\z5}
-2355840 {\z3} {\z4}
\nonumber\\&\hspace{25mm}
-100848{\zeta_{5,3}}
-784620 {\z7}
+38018669 {\z8}
\Bigg)
\nonumber\\&
+\frac{32}{9}\frac{1}{\barom^4} \Bigg(
-38880 {\zeta_3^2}
+68832 {\z2}{\z5}
+269136 {\z3} {\z4}
\nonumber\\&\hspace{25mm}
+10784{\zeta_{5,3}}
+32220 {\z7}
-4254091{\z8}
\Bigg)
\nonumber\\&
+\frac{64}{45}\frac{1}{\barom^5} \Bigg(
23040 {\zeta_3^2}
-23040 {\z2} {\z5}\
-155520 {\z3} {\z4}
\nonumber\\&\hspace{25mm}
-5536 {\zeta_{5,3}}
-5760 {\z7}
+2275973 {\z8}
\Bigg)
\Bigg]\ \ +\ \ \cdots\,.
\end{align}
The most singular  term in $\barom$, proportional to $g^{2\ell}/\barom^\ell$, has a rather simple form, at least for the first three orders. The term proportional to $g^4/\barom$ can be generated with the following expression:
\begin{equation}
\sum_{n=1}\frac{1+n}{2^n}\,(-\omega)^n\,\zeta_{n+2}\,.
\end{equation}
The other terms are much more complicated and we did not find any generated expressions for them with the available data.

The solution of Eq.~(\ref{DLgenerC}) has a very simple form
\begin{equation}\label{DLgenerCSolution}
\gamma\,=-\,\omega+\sqrt{\ \omega+\sum_{k=1}\sum_{m=0}{\mathfrak C}_m^k\,\omega^m\,g^{2k}}
\end{equation}
and provides the results for the definite contributions into analytically continued anomalous dimension of twist-2 operators in {\emph{all orders}} of perturbative theory after expansion over $g^2$. For example, for the contribution that does not contain any $\zeta_i$, we have from
Eq.~(\ref{DLgener1})
\begin{eqnarray}
\gamma&=&
8\,g^2\left(
-\frac{1}{\omega}
+1
+\omega
+\omega^2
+\omega^3
+\cdots
\right)
+32\,g^4\left(
-\frac{1}{\omega^3}
+\frac{2}{\omega^2}
+\frac{1}{\omega}
-\omega
-2 \omega^2
-3 \omega^3
+\cdots
\right)\nonumber\\&&
+256\,g^6\left(
-\frac{1}{\omega^5}
+\frac{3}{\omega^4}
-\frac{2}{\omega^2}
-\frac{3}{\omega}
-3
-2  \omega
+3 \omega^3
+\cdots
\right)\ +\ \cdots\,.
\end{eqnarray}
Note that we can predict not only the poles, but also the regular part of $\omega$-expansion\footnote{However, the results for the regular part don't give any new information for the reconstruction of the anomalous dimension from the constraints, coming from the generalised double-logarithmic equation, see Appendix~\ref{App:ReconAD}.}.
Similar expansions can be easily obtained for other $\zeta_i$ contributions.

\subsection{Analytical continuation from odd values of $M$}\label{SubSection:3:1}

Having in hand the analytical continuation for all harmonic sums up to weight (transcendentality level) $12$ not only from even values of $\M$, but also from the odd values of $\M$~\cite{Velizhanin:2020avm}, we studied the properties of the analytical continuation of anomalous dimension from the odd positive values $\M$. Actually, the anomalous dimension of all multiplicatively-renormalised twist-2 operators in $\cN=4$ SYM theory is expressed through one universal anomalous dimension with the shifted arguments, which has {\it only even} argument. This can be seen from the explicit results at one~\cite{Kotikov:2002ab} and two~\cite{Kotikov:2003fb} loops and was shown in the general case in Ref.~\cite{Belitsky:2003sh}, smilar to quasi-partonic operators in QCD~\cite{Bukhvostov:1985rn}. 
However, if we take the expression for the planar seven-loop anomalous dimension of twist-2 operators and perform the analytical continuation not from even, but from {\em{odd}} values of~$\M$ near~$\M=-1+\omega$, we obtain the equation, which is very close to the generalised double-logarithmic equation~(\ref{DLgener})\footnote{The math-file with full result can be found in the ancillary files of the arXiv version of this paper.}:
\begin{eqnarray}\label{DLOddgener}
\gamma\,(2\,\omega+\gamma)&=&16\,g^2 \Big[-1+\z2\,\omega ^2-\z3\,\omega ^3+\z4\,\omega ^4-\z5\, \omega ^5+\z6\, \omega^6-\z7\, \omega^7\Big]\nonumber\\[2mm]&&
+\,g^4 \bigg[
-64\, \z2
+\omega  (+96\, \z3)
+\,\omega ^2 (-8\, \z4)
+\,\omega ^3 (-256\, \z2\z3+360\, \z5)\nonumber\\&&\qquad\
+\,\omega ^4 \left(144\, \zeta_3^2+\frac{58}{3} \z6\right)
+\,\omega ^5 \big(-384\, \z2\z5-280\, \z4\z3+707\, \z7\big)
\bigg]\nonumber\\[2mm]&&
+\,g^6 \bigg[
256\, \z4
+\omega  (1152\, \z2\z3-960\, \z5)
+\,\omega ^2 \left(-1056\, \zeta_3^2-\frac{5000}{3} \z6\right)\nonumber\\&&\qquad\
+\,\omega ^3 \bigg(6880\, \z2 \z5-1072\, \z4 \z3-6412\, \z7\bigg)
\bigg]\nonumber\\[2mm]&&
+\,g^8 \bigg[
384\, \zeta_3^2+\frac{1888}{3} \z6+16\,\omega  \left(-1296\, \z2 \z5+452\,\z4 \z3+553\, \z7\right)
\bigg],\qquad\quad
\end{eqnarray}
where we write the right-hand side only up to the fourth order and up to transcedentality~$7$, while the full results up to seven loops and up to transcedentality $12$ can be found in Appendix~\ref{DLoddF}.
In fact, the differences come only with the terms, which are proportional to the product of the special numbers $\zeta_i$ and harmonic sums with the corresponding argument: for the usual double-logarithmic equation~(\ref{DLgener}) the argument of harmonic sums is equal to~$1$ and its contribution to the result is the same (up to sign) for all harmonic sums as $S_{i_1,i_2,...,i_k}(1)=\pm 1$, while for the analytical continuation from odd positive values of $\M$ to negative $\M=-1+\omega$ all harmonic sums are equal to~$0$ as $S_{i_1,i_2,...,i_k}(0)=0$. 

Note that $\M=-1+\omega$ is the value related with the BFKL equation. The difference between analytical continuation from the even and from the odd positive integer values comes from the sign, which appear in some terms for the analytical continuation of the harmonic sums with negative indices (see Eq.~(\ref{Sm21})). With plus sign we obtain the BFKL single-logarithms, while for minus sign we obtain the double-logarithms, which combine in the form of Eq.~(\ref{DLOddgener}).

\setcounter{footnote}{0}
\section{Conclusion}

We studied the properties of the generalised double-logarithmic equation~\cite{Velizhanin:2011pb} for the seven-loop planar anomalous dimension of twist-2 operators in $\cN=4$ SYM theory, analytically continued into $\M=-2+\omega$ up to transcedental numbers with weight $12$ ($\zz12$ and similar), proposed in Ref.~\cite{Velizhanin:2011pb} and found that it can be resummed into the inverse powers of $\barom=(1-\omega)$. The resummed expression~(\ref{DLgenerRessumALL}) provides much more information about $\omega$-expansion of anomalous dimension in any-loop order. Moreover, we have found the similar generalised double-logarithmic equation, but for the anomalous dimension of twist-2 operators in the planar $\cN=4$ SYM theory, analytically continued from {\emph{odd}} positive integers values to $\M=-1+\omega$, assuming that the anomalous dimension has the same form for even and for odd positive integer values of argument. Both equations provide much more information about the anomalous dimension of twist-2 operators compared to the information obtained from the BFKL equation\footnote{The study of the relation between BFKL and DGLAP equations performed in Ref.~\cite{Kotikov:2002ab} provides much more information from the BFKL equation for the anomalous dimension of twist-2 operators, but it is not clear how to generalised this approach to higher orders and, to our knowledge, the corresponding results were not used for the reconstructions of the anomalous dimension of twist-2 operators.}. In principle, such information is sufficient to reconstruct the anomalous dimension of twist-2 operators at any order (see Appendix~\ref{App:ReconAD}) using \texttt{LLL}-algorithm~\cite{Lenstra1982} from the number theory  (see Refs.~\cite{Velizhanin:2011pb,Velizhanin:2013vla}), if the analytical continuation for the relevant harmonic sums is known. However, in practice there is a limitation related with the applicability of the current realisation of \texttt{LLL}-algorithm\footnote{We use \texttt{fplll}-labriry~\cite{fplll} for these purposes.} for huge matrices.

The generalised double-logarithmic equation is one of the few results, which has been obtained in $\cN=4$ SYM theory, but it works perfectly in realistic QCD for the non-singlet anomalous dimension of twist-2 operators if we add $\beta$-function as~\cite{Velizhanin:2014dia}\footnote{There is a difference between the normalisation of the anomalous dimension in $\cN = 4$ SYM theory
and in QCD, which produces the difference in the left-hand sides of Eqs.~(\ref{DLgenerALL}) and~(\ref{GDLQCD})}:
\begin{equation}\label{GDLQCD}
\gamma_{\mathrm{NS}}\Big(\omega+\gamma_{\mathrm{NS}}-\beta/\alpha_s\Big)={\mathcal{O}}\left(\omega^0\right)+{\mathrm{poles~terms}}\,,
\end{equation}
where the poles terms appear for the first time in three-loop order and are proportional to $\z2$ only, moreover, they disappear completely in the planar limit of QCD, when the Casimir operators for the fundamental and adjoint representations of $SU(n_c)$ gauge group are reduced to $C_F=n_c/2$ and $C_A=n_c$.

Note also, that there is the general expression for the non-planar contribution to the four-loop universal anomalous dimension in $\cN=4$ SYM theory~\cite{Kniehl:2021ysp}, which was reconstructed for arbitrary $\M$ from the result for the fixed values~\cite{Velizhanin:2009gv,Velizhanin:2010ey,Velizhanin:2014zla,Kniehl:2020rip} (see also Ref.~\cite{Fleury:2019ydf}). The double-logarithmic limit $M=-2+\omega$ for this result can be written as~\cite{Kniehl:2021ysp}:
\begin{eqnarray}
\gamma(-2+\omega)&=
&\gamma_{\mathrm {planar}}+\frac{48}{N_c^2}\,g^8\Bigg[
\frac{192 {\z2}}{\omega^5}
-\frac{384 {\z2}}{\omega^4}
-\frac{48}{\omega^3}\left(4 {\z2}+15 {\z4}\right)
+\frac{288}{\omega^2}\left(4 {\z2} {\z3}+5 {\z4}\right)\nonumber\\&&\hspace*{-10mm}
{}+\frac{4 }{\omega}
\left(
144 {\z2}
-1728 {\z2} {\z3}
-24 {\z3}^2
-540 {\z4}
+60 {\z5}
-1367 {\z6}
\right)\nonumber\\&&\hspace*{-10mm}
{}+4 \left(
96 {\z2}
-288 {\z2} {\z3}
+722 {\z2} {\z5}
+36 {\z3}^2
-814 {\z3} {\z4}
+799 {\z6}
-56 {\z7}
\right)+\mathcal{O}(\omega)\bigg]\qquad\label{DLlimit}
\end{eqnarray}
and it violates our generalised double-logarithmic equation~(\ref{DLgenerC}).

At this moment it is not clear how to obtain the corrections to the double-logarithmic equation~(\ref{DL}) directly from the diagrammatic calculations or by means of QCS-approach. However, its simplicity and informativeness are very attractive.

\acknowledgments

I would like to thank L. Lipatov, A. Onishchenko, M. Ryskin and A. Shuvaev for useful discussions.
This research was supported by by RFBR grants 19-02-00983-a and 16-02-00943-a and a Marie Curie International Incoming Fellowship within the 7th European Community Framework Programme, grant number PIIF-GA-2012-331484.

\appendix
\section{Full generalised double-logarithmic equation near $\M=-2+\omega$}\label{Section:DLgenerALL}

In this Appendix we write down the generalised double-logarithmic equation~(\ref{DLgener}) from all available data. We obtain this equation from the planar seven-loop anomalous dimension~\cite{Kotikov:2002ab,Kotikov:2003fb,Kotikov:2004er,Staudacher:2004tk,Kotikov:2007cy,Bajnok:2008qj,Lukowski:2009ce,Marboe:2014sya,Marboe:2016igj} with our database~\cite{Velizhanin:2020avm} for the analytical continuation of the harmonic sums with maximal weight $13$ up to $\zz12$ and similar multiple zeta values (MZV) with weight $12$. The database for the relations between MZV from ~\cite{Blumlein:2009cf}, which we used for the analytical continuation, contains the basis of MZV consisting of $\zeta_i$ and ${\mathrm h}_{\vec{m}}$, for which
\begin{equation}
{\mathrm h}_{i_1 i_2 i_3 \ldots}=(-1)^k\,{\mathrm{Li}}_{i_1,i_2,i_3,\ldots}(-i_1,i_2,i_3,\ldots)
\end{equation}
where $i_n>0$, $k$ is the length of list $\{i_1,i_2,i_3,\ldots\}$ and $\mathrm{Li}_{\vec{m}}(\vec{x})$ is the multiple polylogarithms as defined in Ref.~\cite{Vollinga:2004sn}. However, we have found, that all ${\mathrm h}_{\vec{m}}$ are combined in the single-valued MZV's, as given in Ref.~\cite{Schnetz:2016fhy} (see also Ref.~\cite{Broadhurst:1995km}), that is in terms of $\zeta_i$, $\zeta_{5,3}$, $\zeta_{7,3}$, $\zeta_{3,5,3}$, $\zeta_{9,3}$ and $\zeta_{4,4,2,2}$. The final result is the following\footnote{The math-file with this result can be found in the ancillary files of the arXiv version of this paper}:
\begin{align}\label{DLgenerALL}
\gamma\,(2\,\omega+\gamma)&=
16\,g^2\Bigg[
  -1
  +\omega
  +\omega^2(
  1
  +\z2)
  +\omega^3(
  1
  -\z3)
  +\omega^4(
  1
  +\z4)
  +\omega^5(
  1
  -\z5)
\nonumber\\&\hspace*{16mm}
  +\omega^6(
  1
  +\z6)
   +\omega^7(
  1
  -\z7)
  +\omega^8(
  1
  +\z8)
   +\omega^9(
  1
  -\z9)
\nonumber\\&\hspace*{16mm}
   +\omega^{10}(
  1
  +\zz10)
   +\omega^{11}(
  1
  -\zz11)
   +\omega^{12}(
  1
  +\zz12)\Bigg]
\nonumber\\&\hspace*{-10mm}
+
g^4\Bigg[
  -64\,\z2
  +\omega\Big(
  128\,\z2
  +96\,\z3\Big)
  +\omega^2\Big(
  192\,\z2
  -160\,\z3
  -8\,\z4\Big)
\nonumber\\&
  +\omega^3\bigg(
  256\,\z2
  -224\,\z3
  -256\,\z2\,\z3
  +152\,\z4
  +360\,\z5\bigg)
\nonumber\\&
  +\omega^4\bigg(
  320\,\z2
  -288\,\z3
  +144\,\zeta_3^2
  +216\,\z4
  -144\,\z5
  +\frac{58}{3}\z6\bigg)
\nonumber\\&
  +\omega^5\bigg(
  384\,\z2
  -352\,\z3
  +280\,\z4
  -280\,\z3\,\z4
  -208\,\z5
  -384\,\z2\,\z5
  +138\,\z6
  +707\,\z7\bigg)
\nonumber\\&
  +\omega^6\bigg(
  448\,\z2
  -416\,\z3
  +344\,\z4
  -272\,\z5
  +272\,\z3\,\z5
  +202\,\z6
  -134\,\z7
  -\frac{144}{5}\zeta_{5,3}
\nonumber\\&
  +\frac{1183}{30}\z8\bigg)
  +\omega^7\bigg(
  512\,\z2
  -480\,\z3
  +408\,\z4
  -336\,\z5
  -408\,\z4\,\z5
  +266\,\z6
\nonumber\\&
  -266\,\z3\,\z6
  -198\,\z7
  -512\,\z2\,\z7
  +1185\,\z9
  +\frac{263}{2}\z8\bigg)
  +\omega^8\bigg(
  576\,\z2
  -544\,\z3
\nonumber\\&
  +472\,\z4
  -400\,\z5
  +330\,\z6
  -262\,\z7
  +262\,\z3\,\z7
  -130\,\z9
  -\frac{141}{7}\zeta_{7,3}
  +\frac{391}{2}\z8
\nonumber\\&
  +\frac{977}{7}\zeta_5^2
  +\frac{12097}{280}\zz10\bigg)
  +\omega^9\bigg(
  640\,\z2
  -608\,\z3
  +536\,\z4
  -464\,\z5
  +394\,\z6
  -326\,\z7
\nonumber\\&
  -394\,\z5\,\z6
  -536\,\z4\,\z7
  -194\,\z9
  -640\,\z2\,\z9
  +\frac{519}{2}\z8
  +\frac{1033}{8}\zz10
  +\frac{28677}{16}\zz11
\nonumber\\&
  -\frac{519}{2}\z3\z8\bigg)
  +\omega^{10}\bigg(
  704\,\z2
  -672\,\z3
  +600\,\z4
  -528\,\z5
  +458\,\z6
  -390\,\z7
  -258\,\z9
\nonumber\\&
  +298\,\z5\,\z7
  +258\,\z3\,\z9
  -\frac{46}{3}\zeta_{9,3}
  +\frac{647}{2}\z8
  -\frac{1029}{8}\zz11
  +\frac{1545}{8}\zz10
  +\frac{1161445}{66336}\zz12\bigg)
\Bigg] 
\nonumber\\&\hspace*{-10mm}
+g^6\Bigg[
  128\,\z3
  +256\,\z4
  +\omega\Big(
  512\,\z3
  +1152\,\z2\,\z3
  +672\,\z4
  -960\,\z5\Big)
\nonumber\\&
  +\omega^2\bigg(
  384\,\z3
  -2688\,\z2\,\z3
  -1056\,\zeta_3^2
  +256\,\z4
  +1504\,\z5
  -\frac{5000}{3}\z6\bigg)
\nonumber\\&
  +\omega^3\bigg(
  -256\,\z3
  -3968\,\z2\,\z3
  +1760\,\zeta_3^2
  +1248\,\z4
  -1072\,\z3\,\z4
  +4224\,\z5
  -6412\,\z7
\nonumber\\&
  +6880\,\z2\,\z5
  +\frac{11696}{3}\z6\bigg)
  +\omega^4\bigg(
  -1408\,\z3
  -5760\,\z2\,\z3
  +2720\,\zeta_3^2
  +3648\,\z4
\nonumber\\&
  +3552\,\z2\,\zeta_3^2
  -1904\,\z3\,\z4
  +5920\,\z5
  -4800\,\z2\,\z5
  -6080\,\z3\,\z5
  +2488\,\z7
  +\frac{5080}{3}\z6
\nonumber\\&
  -\frac{9170}{3}\z8
  +384\,\zeta_{5,3}\bigg)
  +\omega^5\bigg(
  -3072\,\z3
  -8064\,\z2\,\z3
  +3936\,\zeta_3^2
  -768\,\z2\,\zeta_3^2
  -1504\,\zeta_3^3
\nonumber\\&
  +7456\,\z4
  -4048\,\z3\,\z4
  +6592\,\z5
  -5952\,\z2\,\z5
  +1824\,\z3\,\z5
  -968\,\z4\,\z5
  +1664\,\z6
\nonumber\\&
  +9012\,\z7
  +18648\,\z2\,\z7
  -\frac{4032}{5}\zeta_{5,3}
  -\frac{187304}{9}\z9
  +\frac{347116}{45}\z8
  -\frac{20704}{3}\z3\z6\bigg)
\nonumber\\&
  +\omega^6\bigg(
  -5248\,\z3
  -10880\,\z2\,\z3
  +5408\,\zeta_3^2
  -768\,\z2\,\zeta_3^2
  +256\,\zeta_3^3
  +12672\,\z4
  +6240\,\z5
\nonumber\\&
  -6704\,\z3\,\z4
  +3856\,\zeta_3^2\,\z4
  -7616\,\z2\,\z5
  +4192\,\z3\,\z5
  +10976\,\z2\,\z3\,\z5
  -4184\,\z4\,\z5
\nonumber\\&
  -4496\,\zeta_5^2
  +14000\,\z7
  -6776\,\z2\,\z7
  -10332\,\z3\,\z7
  -\frac{3456}{5}\zeta_{5,3}
  +\frac{11432}{3}\z6
  +\frac{38756}{9}\z9
\nonumber\\&
  +\frac{75718}{45}\z8
  -\frac{9363}{5}\zz10
  +\frac{12628}{3}\z3\z6
  -576\,\z2\,\zeta_{5,3}
  +648\,\zeta_{7,3}\bigg)
\nonumber\\&
  +\omega^7\bigg(
  4864\,\z5
  -7936\,\z3
  -14208\,\z2\,\z3
  +7136\,\zeta_3^2
  -768\,\z2\,\zeta_3^2
  +256\,\zeta_3^3
  +256\,\z2\,\zeta_3^3
\nonumber\\&
  +19296\,\z4
  -9872\,\z3\,\z4
  -768\,\zeta_3^2\,\z4
  -9792\,\z2\,\z5
  +7072\,\z3\,\z5
  -3840\,\z2\,\z3\,\z5
\nonumber\\&
  +17452\,\z7
  -3968\,\zeta_3^2\,\z5
  -6200\,\z4\,\z5
  -12944\,\z5\,\z6
  -7848\,\z2\,\z7
  -2084\,\z3\,\z7
\nonumber\\&
  +1533\,\z4\,\z7
  +\frac{576}{5}\zeta_{3,5,3}
  -\frac{1060}{7}\zeta_5^2
  -\frac{5076}{7}\zeta_{7,3}
  -\frac{12640}{9}\z8
  +\frac{24400}{3}\z6
\nonumber\\&
  +\frac{145352}{9}\z9
  -\frac{3095713}{60}\zz11
  +\frac{6849011}{525}\zz10
  -\frac{2304}{5}\z2\zeta_{5,3}
  +\frac{4608}{5}\z3\zeta_{5,3}
\nonumber\\&
  +\frac{4828}{3}\z3\z6
  +\frac{357092}{9}\z2\z9
  -\frac{615577}{45}\z3\z8
  -576\,\zeta_{5,3}\bigg)
\nonumber\\&
  +\omega^8\bigg(
  -11136\,\z3
  -18048\,\z2\,\z3
  +9120\,\zeta_3^2
  -768\,\z2\,\zeta_3^2
  +256\,\zeta_3^3
  -128\,\zeta_3^4
\nonumber\\&
  +27328\,\z4
  -13552\,\z3\,\z4
  -768\,\zeta_3^2\,\z4
  +2464\,\z5
  -12480\,\z2\,\z5
  +10464\,\z3\,\z5
\nonumber\\&
  -3840\,\z2\,\z3\,\z5
  +768\,\zeta_3^2\,\z5
  -8728\,\z4\,\z5
  +18472\,\z3\,\z4\,\z5
  +14632\,\z6
  +19368\,\z7
\nonumber\\&
  -9432\,\z2\,\z7
  +756\,\z3\,\z7
  +9348\,\z2\,\z3\,\z7
  -6337\,\z4\,\z7
  -7294\,\z5\,\z7
  -8760\,\z2\,\z9
\nonumber\\&
  -\frac{2304}{5}\zeta_{5,3}
  -\frac{2304}{5}\zeta_{3,5,3}
  +\frac{2512}{3}\zeta_{9,3}
  -\frac{4512}{7}\zeta_{7,3}
  +\frac{8472}{7}\zeta_5^2
  -\frac{69638}{45}\z8
\nonumber\\&
  +\frac{99763}{15}\zz11
  +\frac{233516}{9}\z9
  +\frac{1780717}{1050}\zz10
  +\frac{449442803}{82920}\zz12
  +\frac{2112}{5}\z4\zeta_{5,3}
\nonumber\\&
  -\frac{2304}{5}\z2\zeta_{5,3}
  -\frac{4508}{3}\z3\z6
  +\frac{4876}{3}\zeta_3^2\z6
  -\frac{5484}{7}\z2\zeta_{7,3}
  +\frac{6082}{3}\z5\z6
\nonumber\\&
  +\frac{39828}{7}\z2\zeta_5^2
  -\frac{143216}{9}\z3\z9
  +\frac{224179}{15}\z3\z8
  +192\,\zeta_{4,4,2,2}
\bigg)
\Bigg]
\nonumber\\&\hspace*{-10mm}
+g^8\Bigg[
  2560\,\z2\,\z3
  +384\,\zeta_3^2
  -128\,\z5
  +\frac{1888}{3}\z6
  +\omega\bigg(
  4608\,\z2\,\z3
  -2944\,\zeta_3^2
\nonumber\\&
  +7232\,\z3\,\z4
  -10880\,\z5
  -20736\,\z2\,\z5
  +8848\,\z7
  -\frac{70624}{3}\z6\bigg)
  +\omega^2\bigg(
  3072\,\z2\,\z3
\nonumber\\&
  -7424\,\zeta_3^2
  -20224\,\z2\,\zeta_3^2
  -35136\,\z3\,\z4
  -8704\,\z5
  +58880\,\z2\,\z5
  +24384\,\z3\,\z5
\nonumber\\&
  -704\,\z7
  -\frac{3968}{5}\zeta_{5,3}
  +\frac{48512}{3}\z6
  +\frac{1646704}{45}\z8\bigg)
  +\omega^3\bigg(
  -8192\,\z2\,\z3
  -5888\,\zeta_3^2
\nonumber\\&
  +48384\,\z2\,\zeta_3^2
  +11712\,\zeta_3^3
  -33408\,\z3\,\z4
  +4352\,\z5
  +70400\,\z2\,\z5
  -29120\,\z3\,\z5
\nonumber\\&
  +84432\,\z4\,\z5
  +58256\,\z3\,\z6
  -100912\,\z7
  -165184\,\z2\,\z7
  +\frac{31616}{5}\zeta_{5,3}
  +\frac{60224}{3}\z6
\nonumber\\&
  +\frac{935056}{9}\z9
  -\frac{4815368}{45}\z8\bigg)
  +\omega^4\bigg(
   7808\,\zeta_3^2
  -35328\,\z2\,\z3
  +66304\,\z2\,\zeta_3^2
  -22464\,\zeta_3^3
\nonumber\\&
  -57216\,\z3\,\z4
  +47344\,\zeta_3^2\,\z4
  +26240\,\z5
  +94720\,\z2\,\z5
  -101952\,\z3\,\z5
  -132096\,\z7
\nonumber\\&
  -204224\,\z2\,\z3\,\z5
  -123152\,\z4\,\z5
  -203312\,\z3\,\z6
  +194880\,\z2\,\z7
  +108768\,\z3\,\z7
\nonumber\\&
  -\frac{10624}{5}\zeta_{5,3}
  -\frac{32456}{7}\zeta_{7,3}
  +\frac{41696}{3}\z6
  +\frac{138544}{9}\z9
  +\frac{307512}{7}\zeta_5^2
  +\frac{2756672}{45}\z8
\nonumber\\&
  +\frac{34103276}{525}\zz10
  +\frac{16256}{5}\z2\zeta_{5,3}
\bigg)
  +\omega^5\,(
   39808\,\zeta_3^2
  -84480\,\z2\,\z3
\nonumber\\&
  +94976\,\z2\,\zeta_3^2
  -32960\,\zeta_3^3
  -50432\,\z2\,\zeta_3^3
  -129600\,\z3\,\z4
  +6128\,\zeta_3^2\,\z4
  +54912\,\z5
\nonumber\\&
  +119552\,\z2\,\z5
  -167744\,\z3\,\z5
  +228992\,\z2\,\z3\,\z5
  +88848\,\zeta_3^2\,\z5
  -96112\,\z4\,\z5
\nonumber\\&
  -103472\,\z7
  +245056\,\z2\,\z7
  +24528\,\z3\,\z7
  +331188\,\z4\,\z7
  +506717\,\zz11
  +\frac{70112}{3}\z6
\nonumber\\&
  -\frac{41344}{5}\zeta_{5,3}
  +\frac{104840}{7}\zeta_{7,3}
  +\frac{120088}{7}\zeta_5^2
  -\frac{3098608}{9}\z9
  +\frac{3897592}{45}\z8
  -\frac{260384}{3}\z3\z6
\nonumber\\&
  -\frac{35143176}{175}\zz10
  -\frac{31488}{5}\z2\zeta_{5,3}
  -\frac{58592}{5}\z3\zeta_{5,3}
  +\frac{757764}{5}\z3\z8
  +\frac{765064}{3}\z5\z6
\nonumber\\&
  -\frac{1966496}{3}\z2\z9
  -2976\,\zeta_{3,5,3}\bigg)
  +\omega^6\bigg(
  -161792\,\z2\,\z3
  +96256\,\zeta_3^2
\nonumber\\&
  +137472\,\z2\,\zeta_3^2
  -48576\,\zeta_3^3
  +24064\,\z2\,\zeta_3^3
  +17784\,\zeta_3^4
  -273600\,\z3\,\z4
  +25520\,\zeta_3^2\,\z4
\nonumber\\&
  +88320\,\z5
  +132608\,\z2\,\z5
  -208064\,\z3\,\z5
  -75440\,\z4\,\z5
  -24256\,\z7
  -545820\,\z5\,\z6
\nonumber\\&
  +213632\,\z2\,\z3\,\z5
  +97216\,\z3\,\z4\,\z5
  -39136\,\zeta_3^2\,\z5
  +334656\,\z2\,\z7
  -465784\,\z2\,\z3\,\z7
\nonumber\\&
  -214976\,\z3\,\z7
  -240456\,\z4\,\z7
  +\frac{30584}{7}\zeta_{7,3}
  -\frac{60544}{5}\zeta_{5,3}
  +\frac{104044}{3}\zz11
  +\frac{222656}{3}\z6
\nonumber\\&
  -\frac{112996}{9}\zeta_{9,3}
  -\frac{477432}{7}\zeta_5^2
  +\frac{1291552}{45}\z8
  -\frac{1538848}{3}\z9
  -\frac{47264}{3}\z3\z6
  +\frac{324910}{3}\z5\z7
\nonumber\\&
  +\frac{119129786}{525}\zz10
  -\frac{81288}{5}\z4\zeta_{5,3}
  -\frac{86784}{5}\z2\zeta_{5,3}
  +\frac{100344}{7}\z2\zeta_{7,3}
  +\frac{125568}{5}\z3\zeta_{5,3}
\nonumber\\&
  +\frac{599780}{3}\zeta_3^2\z6
  -\frac{1486152}{7}\z2\zeta_5^2
  +\frac{2520308}{9}\z3\z9
  +\frac{4045568}{9}\z2\z9
  -\frac{24804844}{45}\z3\z8
\nonumber\\&
  +\frac{37008793}{13820}\zz12
  +13056\,\zeta_{3,5,3}
  -2856\,\zeta_{4,4,2,2}\bigg)
\Bigg]
\nonumber\\&\hspace*{-10mm}
+g^{10}\Bigg[
  7168\,\z2\,\zeta_3^2
  +19456\,\z3\,\z4
  -37888\,\z2\,\z5
  -3072\,\z3\,\z5
  -11520\,\z7
  +\frac{5376}{5}\zeta_{5,3}
\nonumber\\&
  -\frac{1104928}{45}\z8
  +\omega\bigg(
  7168\,\zeta_3^2
  -78848\,\z2\,\zeta_3^2
  -9984\,\zeta_3^3
  +19456\,\z3\,\z4
  -100352\,\z2\,\z5
\nonumber\\&
  +10752\,\z3\,\z5
  -200896\,\z4\,\z5
  +189824\,\z7
  +335872\,\z2\,\z7
  -\frac{45568}{5}\zeta_{5,3}
  -\frac{216224}{3}\z9
\nonumber\\&
  +\frac{16089344}{45}\z8
  -\frac{309248}{3}\z3\,\z6\bigg)
  +\omega^2\bigg(
  -18432\,\zeta_3^2
  -110592\,\z2\,\zeta_3^2
  +55808\,\zeta_3^3
\nonumber\\&
  -112128\,\z3\,\z4
  -300736\,\zeta_3^2\,\z4
  -71680\,\z2\,\z5
  +286464\,\z3\,\z5
  +754688\,\z2\,\z3\,\z5
\nonumber\\&
  +1074944\,\z4\,\z5
  +127744\,\z7
  -1082880\,\z2\,\z7
  -235680\,\z3\,\z7
  -\frac{357051032}{525}\zz10
\nonumber\\&
  +\frac{74608}{7}\zeta_{7,3}
  +\frac{80128}{5}\zeta_{5,3}
  -\frac{752816}{7}\zeta_5^2
  -\frac{3180160}{9}\z9
  -\frac{15323744}{45}\z8
  +\frac{48128}{5}\z2\,\zeta_{5,3}
\nonumber\\&
  +\frac{1976960}{3}\z3\,\z6
\bigg)
  +\omega^3\bigg(
  -83968\,\zeta_3^2
  -19456\,\z2\,\zeta_3^2
  +99840\,\zeta_3^3
\nonumber\\&
  +323072\,\z2\,\zeta_3^3
  -303104\,\z3\,\z4
  +1131904\,\zeta_3^2\,\z4
  +161792\,\z2\,\z5
  +372992\,\z3\,\z5
\nonumber\\&
  -2162176\,\z2\,\z3\,\z5
  -390272\,\zeta_3^2\,\z5
  +779008\,\z4\,\z5
  -115968\,\z7
  -1431936\,\z2\,\z7
\nonumber\\&
  -235712\,\z3\,\z7
  -2974304\,\z4\,\z7
  +\frac{129024}{5}\zeta_{3,5,3}
  +\frac{532256}{9}\z8
  +\frac{1053254848}{525}\zz10
\nonumber\\&
  -\frac{7598556}{5}\zz11
  -\frac{533696}{7}\zeta_{7,3}
  -\frac{608768}{7}\zeta_5^2
  +\frac{22000864}{9}\z9
\nonumber\\&
  -\frac{1458176}{3}\z3\,\z6
  -\frac{3105200}{3}\z5\,\z6
  +\frac{86528}{5}\z2\,\zeta_{5,3}
  +\frac{174848}{5}\z3\,\zeta_{5,3}
  +\frac{33289088}{9}\z2\,\z9
\nonumber\\&
  -\frac{38284336}{45}\z3\,\z8
  +14848\,\zeta_{5,3}\bigg)
  +\omega^4\bigg(
  -163840\,\zeta_3^2
  +363520\,\z2\,\zeta_3^2
  +72448\,\zeta_3^3
\nonumber\\&
  -840192\,\z2\,\zeta_3^3
  -702464\,\z3\,\z4
  +1199104\,\zeta_3^2\,\z4
  +680960\,\z2\,\z5
  -148992\,\z3\,\z5
\nonumber\\&
  -2257920\,\z2\,\z3\,\z5
  +503808\,\zeta_3^2\,\z5
  +266944\,\z4\,\z5
  -4640256\,\z3\,\z4\,\z5
  -467712\,\z7
\nonumber\\&
  -1293616\,\zeta_3^2\,\z6
  -1769600\,\z2\,\z7
  +2618528\,\z3\,\z7
  +7125264\,\z4\,\z7
  -892496\,\z5\,\z7
\nonumber\\&
  +5041344\,\z2\,\z3\,\z7
  -\frac{17152}{5}\zeta_{5,3}
  +\frac{374448}{7}\zeta_{7,3}
  -\frac{419840}{3}\zeta_3^4
  -\frac{592896}{5}\zeta_{3,5,3}
\nonumber\\&
  +\frac{6855504}{7}\zeta_5^2
  -\frac{12088236}{5}\zz11
  +\frac{22387616}{45}\z8
  +\frac{23909312}{9}\z9
  -\frac{1042930424}{525}\zz10
\nonumber\\&
  +\frac{33728}{5}\z4\,\zeta_{5,3}
  +\frac{55296}{5}\z2\,\zeta_{5,3}
  -\frac{49546816}{9}\z2\,\z9
  +\frac{149042608}{45}\z3\,\z8
\nonumber\\&
  -\frac{42084279598}{31095}\zz12
  +\frac{13350784}{3}\z5\,\z6
  +\frac{15677344}{7}\z2\,\zeta_5^2
  -\frac{15822496}{9}\z3\,\z9
\nonumber\\&
  -\frac{292832}{7}\z2\,\zeta_{7,3}
  -\frac{3929216}{3}\z3\,\z6
  -205312\,\z3\,\zeta_{5,3}
  +65632\,\zeta_{9,3}
  +3968\,\zeta_{4,4,2,2}\bigg)
\Bigg]
\nonumber\\&\hspace*{-10mm}
+g^{12}\Bigg[
  -12288\,\z2\,\zeta_3^2
  -13312\,\zeta_3^3
  +96768\,\zeta_3^2\,\z4
  -4096\,\z3\,\z5
  -148480\,\z2\,\z3\,\z5
\nonumber\\&
  -464384\,\z4\,\z5
  +486400\,\z2\,\z7
  +4096\,\z3\,\z7
  -\frac{22720}{7}\zeta_5^2
  -\frac{52672}{7}\zeta_{7,3}
\nonumber\\&
  +\frac{2388992}{9}\z9
  +\frac{5404288}{21}\zz10
  -\frac{384512}{3}\z3\,\z6
  +10240\,\z2\,\zeta_{5,3}
\nonumber\\&
  +\omega\bigg(
  -45056\,\z2\,\zeta_3^2
  +8192\,\zeta_3^3
  -267264\,\z2\,\zeta_3^3
  -1685248\,\zeta_3^2\,\z4
  -451584\,\z3\,\z5
\nonumber\\&
  +2016256\,\z2\,\z3\,\z5
  +131456\,\zeta_3^2\,\z5
  -629760\,\z4\,\z5
  +2025984\,\z2\,\z7
  +513152\,\z3\,\z7
\nonumber\\&
  +5058656\,\z4\,\z7
  +\frac{552768}{7}\zeta_{7,3}
  +\frac{1031216}{3}\zz11
  +\frac{2188736}{7}\zeta_5^2
  -\frac{27613312}{9}\z9
\nonumber\\&
  -\frac{216192}{5}\z3\,\zeta_{5,3}
  -\frac{237568}{5}\z2\,\zeta_{5,3}
  -\frac{430336}{3}\z3\,\z6
  +\frac{2505344}{3}\z5\,\z6
\nonumber\\&
  +\frac{11939072}{15}\z3\,\z8
  -\frac{2182031648}{525}\zz10
  -\frac{15587072}{3}\z2\,\z9
  -34176\,\zeta_{3,5,3}\bigg)
\nonumber\\&
  +\omega^2\bigg(
  -503808\,\z2\,\zeta_3^2
  -70656\,\zeta_3^3
  +2073600\,\z2\,\zeta_3^3
  -842752\,\zeta_3^2\,\z4
  +601088\,\z3\,\z5
\nonumber\\&
  +4286464\,\z2\,\z3\,\z5
  -179456\,\zeta_3^2\,\z5
  +2674432\,\z4\,\z5
  -2344640\,\zeta_5^2
  -5145280\,\z5\,\z6
\nonumber\\&
  +900096\,\z2\,\z7
  -5350656\,\z3\,\z7
  -28707584\,\z4\,\z7
  +1893440\,\z5\,\z7
  +2173824\,\zeta_3^2\,\z6
\nonumber\\&
  +15663712\,\z3\,\z4\,\z5
  -13296896\,\z2\,\z3\,\z7
  +418048\,\z3\,\zeta_{5,3}
  -178880\,\zeta_{7,3}
\nonumber\\&
  -130608\,\zeta_{9,3}
  +17248\,\zeta_{4,4,2,2}
  +\frac{563552}{3}\zeta_3^4
  +\frac{1330944}{5}\zeta_{3,5,3}
  -\frac{15037696}{9}\z9
\nonumber\\&
  -\frac{41665760}{7}\z2\,\zeta_5^2
  +\frac{168491792}{15}\zz11
  +\frac{309691704914}{31095}\zz12
  +\frac{906336}{5}\z4\,\zeta_{5,3}
\nonumber\\&
  -\frac{388704}{7}\z2\,\zeta_{7,3}
  +\frac{52625792}{3}\z2\,\z9
  +\frac{436224}{5}\z2\,\zeta_{5,3}
  -\frac{79257344}{15}\z3\,\z8
\nonumber\\&
  +\frac{127459744}{25}\zz10
  +\frac{4397312}{3}\z3\,\z6
  +\frac{5426096}{3}\z3\,\z9
  \bigg)
\Bigg]
\nonumber\\&\hspace*{-10mm}
+g^{14}\Bigg[
  -8192\,\zeta_3^3
  -471040\,\z2\,\zeta_3^3
  -26624\,\zeta_3^4
  -790528\,\zeta_3^2\,\z4
  +286720\,\z2\,\z3\,\z5
\nonumber\\&
  +11264\,\zeta_3^2\,\z5
  -2988544\,\z3\,\z4\,\z5
  +4096\,\zeta_5^2
  -16896\,\z5\,\z6
  +1220096\,\z2\,\z3\,\z7
\nonumber\\&
  +98304\,\z3\,\z7
  +10045952\,\z4\,\z7
  -4714304\,\zz11
  -43008\,\zeta_{3,5,3}
  +7680\,\zeta_{4,4,2,2}
\nonumber\\&
  +\frac{559360}{9}\zeta_{9,3}
  -\frac{407552}{5}\z3\zeta_{5,3}
  +\frac{457216}{3}\z3\z9
  -\frac{746496}{7}\z2\zeta_{7,3}
  +\frac{771584}{5}\z4\zeta_{5,3}
\nonumber\\&
  -\frac{47719021024}{31095}\zz12
  -\frac{823808}{3}\zeta_3^2\z6
  +\frac{1607936}{3}\z5\z7
  -\frac{14146304}{45}\z3\z8
\nonumber\\&
  -\frac{54290432}{9}\z2\z9
  +\frac{4011008}{7}\z2\zeta_5^2
\Bigg]\,.
\end{align}

\setcounter{footnote}{0}

\section{Reconstructon of anomalous dimenson from GDLE}\label{App:ReconAD}
The generalised double-logarithmic equation provides information, which can be used for the reconstruction of the anomalous dimension of twist-2 operators. If we know the anomalous dimension at $\ell$-loop order, we can use the generalised double-logarithmic equation to obtain the constraints for the $\omega$-expansion of the anomalous dimension at $(\ell+1)$-loop orders. Each harmonic sum can be uniquely identified through its pole structure (see Section~\ref{Section:1}), but since we are restricted only with value $\M=-2+\omega$, we lose some information.
To reconstruct the anomalous dimension of twist-2 operators in second (or two-loop) order, we should, following the maximal-transcedentality principle~\cite{Kotikov:2002ab}, take the basis from all harmonic sums with weight $3$:
\begin{equation}
{\mathrm{B}}_{\mathrm{2-loop}}=\left\{S_{-3},S_3,S_{-2,1},S_{2,1},S_{1,-2},S_{1,2},S_{1,1,1}\right\}\,.\label{BasisL2}
\end{equation}
The analytical continuations for these harmonic sums in $\M=-2+\omega$ look like:
\begin{eqnarray}
S_{-3}(-2+\omega)&=&
- \frac{1}{\omega^3}
- 1 
+ \omega \left(-3  - \frac{21}{8}\z4 \right) 
+ \omega^2 \left(-6 + \frac{45}{8}\z5 \right) 
, \nonumber\\
S_{3}(-2+\omega)&=&
 1 
- \frac{1}{\omega^3}
+ \omega (3 + 3 \z4) 
+ \omega^2 (6 - 6 \z5) 
, \nonumber\\
S_{-2,1}(-2+\omega)&=&
\frac{1}{\omega}(-1 - \z2)
- 2 
+ \z3 
+ \omega \left(-3 + \z2 - \frac{33}{16}\z4 \right) \nonumber\\&&\quad
+\ \omega^2 \left(-4 + 2 \z2 - \z3 - \frac{3}{2}\z2 \z3  + \frac{83}{16}\z5 \right) 
, \nonumber\\
S_{2,1}(-2+\omega)&=&
 2 
+ \frac{1}{\omega}(1 - \z2)
+ \z3 
+ \omega \left(3 - \z2 + \frac{3}{4} \z4\right)  \nonumber\\&&\quad
+\ \omega^2 \left(4 - 2 \z2 + \z3 + 2 \z2 \z3 - \frac{9}{2}\z5 \right) 
, \nonumber\\
S_{1,-2}(-2+\omega)&=&
- \frac{1}{\omega^2}
- \frac{1}{\omega}
- 1 
+  \frac{3}{2} \z3
+ \omega \left(-1 - \frac{3}{2}\z3  - \frac{67}{16}\z4 \right)  \nonumber\\&&\quad
+\ \omega^2 \left(-1 -  \frac{3}{2}\z3 +  \frac{21}{8}\z4 + \frac{83}{16}\z5 \right) 
, \nonumber\\
S_{1,2}(-2+\omega)&=&
- \frac{1}{\omega^2}
- \frac{1}{\omega}
- 1 
- 2 \z3 
+ \omega \left(-1 + 2 \z3 +  \frac{17}{4}\z4\right)  \nonumber\\&&\quad
+\ \omega^2 \left(-1 + 2 \z3 - 3 \z4 -  \frac{9}{2}\z5\right) 
, \nonumber\\
S_{1,1,1}(-2+\omega)&=&
-\z2 
- \z3 
+ \omega \left(-\z2 + 2 \z3 +  \frac{5}{4}\z4\right)  \nonumber\\&&\quad
+\ \omega^2 \left(-\z2 + 2 \z3 + 2 \z2 \z3 - 4 \z5 -  \frac{5}{4}\z4\right) .\label{ACBL2}
\end{eqnarray}
From one-loop result~(\ref{LOAD}) and~(\ref{LOADpolep}) we know, according to Table~\ref{Table:DLgener}, the information about all terms in $\omega$-expansion, which don't contain any transcendental numbers $\zeta_i$. From Eq.~(\ref{ACBL2}) we see that $S_{1,-2}$ and $S_{1,2}$ have exactly the same rational part of $\omega$-expansion, $S_{-2,1}$ and $S_{2,1}$ have the opposite sign for the rational part of $\omega$-expansion and $S_{-3}$ and $S_{3}$ have the opposite sign for the regular part of $\omega$-expansion without transcendental numbers $\zeta_i$, while $S_{1,1,1}$ does not contain such expansion at all. This allows us to fix four coefficients from seven in the ansatz from basis~(\ref{BasisL2}), that is three pole terms and one regular term work here, while the coefficient in the front of $S_{1,1,1}$ remains completely unfixed\footnote{However, it does not satisfy large $\M$ asymptotics.}. Nevertheless, the system of four Diophantine equation for seven variables can be solved with the number theory, applied LLL-algorithm~\cite{Lenstra1982} to the matrix constructed from these equations (see Refs.~\cite{Velizhanin:2010cm,Velizhanin:2013vla} for details)\footnote{We use \texttt{fplll}-labriry~\cite{fplll} for this purpose.}. 

Note, that the basis~(\ref{BasisL2}) can be reduced with the help of Gribov-Lipatov reciprocity~\cite{Dokshitzer:2005bf,Dokshitzer:2006nm,Basso:2006nk}, especially in higher orders.
The reciprocity-respecting function 
${\mathcal{P}}(\M)$~\cite{Dokshitzer:2005bf,Dokshitzer:2006nm,Basso:2006nk} is defined as
\begin{equation} \label{Pfunction}
\gamma(M) = {\mathcal{P}} \left(M+\frac{1}{2} \gamma(M) \right)\,
\end{equation}
and is related to the reciprocity-respecting splitting function ${\mathcal P}(x)$~\cite{Dokshitzer:2005bf,Dokshitzer:2006nm} through the Mellin transformation. In all orders of perturbation theory, ${\mathcal P}(x)$ should satisfy the Gribov-Lipatov relation~\cite{Gribov:1972ri}
\begin{equation}\label{GrLipRel}
{\mathcal P}(x)=-\,x\,{\mathcal P}\!\left(\frac{1}{x}\right)\, .
\end{equation}
The advantage of such consideration is that ${\mathcal{P}}(\M)$ 
can only be expressed in terms of the binomial sums with \emph{positive} indices (see \cite{Vermaseren:1998uu})
\begin{equation}\label{BinomialSums}
\HBS_{i_1,\ldots,i_k}(N)=(-1)^N\sum_{j=1}^{N}(-1)^j\binom{N}{j}\binom{N+j}{j}\HS_{i_1,...,i_k}(j)\,,
\end{equation}
which are the special combinations of the usual harmonic sums~\cite{Dokshitzer:2006nm,Beccaria:2007bb}. The basis constructed from the binomial sums consists of $2^{2(\ell-1)}$ at $\ell$-loop order instead of $((1 - \sqrt{2})^k + (1 + \sqrt{2})^k)/2$ with $k=2\ell-1$ for the usual harmonic sums (for example, at seven loops we have 4096 vs. 47321). At the second order the basis from the binomial harmonic sums contains four terms
\begin{eqnarray}
{\mathbb{B}}_{\mathrm{2-loop}}&=&\Big\{
\mathbb{S}_{3}=2S_{-3}-4S_{-2,1},\
\mathbb{S}_{2,1}=S_{3},\
\mathbb{S}_{1,2}=2S_{-3}-4S_{1,-2},\nonumber\\&&\qquad
\mathbb{S}_{1,1,1}=2S_{3}-4S_{1,2}-4S_{2,1}+8S_{1,1,1}
\Big\}\,.
\label{BasisL2BS}
\end{eqnarray}
Their analytical continuations near $\M=-2+\omega$ have the following form:
\begin{eqnarray}
&&\mathbb{S}_{3}=
- \frac{2}{\omega^3}
+ \frac{4}{\omega }(1 + \z2)
+ 6 
- 4 \z3 
+ \omega (6 - 4 \z2 + 3 \z4) 
\nonumber\\&&\qquad\qquad
+ \omega^2 \left(4 - 8 \z2 + 4 \z3 + 6 \z2 \z3 - \frac{19}{2} \z5\right)
,\nonumber\\
&&\mathbb{S}_{1,2}=
- \frac{2}{\omega^3}
+ \frac{4}{\omega^2}
+ \frac{4}{\omega}
+ 2 
- 6 \z3 
+ \omega \left(-2 + 6 \z3 + \frac{23}{2}\z4 \right)
\nonumber\\&&\qquad\qquad
+ \omega^2 \left(-8 + 6 \z3  - \frac{21}{2} \z4- \frac{19}{2}\z5 \right) 
,\nonumber\\
&&\mathbb{S}_{2,1}=
- \frac{2}{\omega^3}
+ 2 
+ \omega (6 + 6 \z4) 
+ \omega^2 (12 - 12 \z5)
\,,\nonumber\\
&&\mathbb{S}_{1,1,1}=
- \frac{2}{\omega^3} 
+ \frac{4}{\omega^2}
+ \frac{4}{\omega} \z2
-2 
- 8 \z2 
- 4 \z3 
+ \omega (-2 - 4 \z2 + 8 \z3 - 4 \z4) 
\nonumber\\&&\qquad\qquad
+ \omega^2 (4 \z3 + 8 \z2 \z3 + 2 \z4 - 8 \z5)\,.
\end{eqnarray}
Now all terms without $\zeta_i$ are different for the different $\mathbb{S}_{\vec{m}}$, but when we fix all poles we automatically fix the regular part, so we have only three equations for four variables, which can be solved with \texttt{LLL}-algorithm.

At third order we should include, according to Table~\ref{Table:DLgener}, some terms with $\z2$ and $\z3$. This gives nine equations for sixteen variables (really for fourteen variables, as two of them don't satisfy the large $\M$ limit).

Proceed in the same way, we can reconstruct the $\ell$-loop anomalous dimension from the result at $(\ell-1)$ loops using the generalised double-logarithmic equation. However, in practice, at least for the moment, there are problems when solving the system of the Diophantine equation with the help of \texttt{LLL}-algorithm for huge matrices (with dimension about one thousand and more).

\section{Full generalised double-logarithmic equation near $\M=-1+\omega$}\label{DLoddF}
The full result for the analytically continued seven-loop anomalous dimension from odd values into $\M=-1+\omega$ up to weight $12$ reads as\footnote{The math-file with this result can be found in the ancillary files of the arXiv version of this paper}:
\begin{align}\label{DLMm1F}
\gamma\,(2\,\omega+\gamma)&=
16\,g^2\Bigg[
  -1
  +\omega^2\,\z2
  -\omega^3\,\z3
  +\omega^4\,\z4
  -\omega^5\,\z5
  +\omega^6\,\z6
  -\omega^7\,\z7
\nonumber\\&
  +\omega^8\,\z8
  -\omega^9\,\z9
  +\omega^{10}\,\zz10
  -\omega^{11}\,\zz11
  +\omega^{12}\,\zz12
  \Bigg]
\nonumber\\&\hspace*{-10mm}
+g^4\Bigg[
  -64\,\z2
  +96\,\omega\,\z3
  -8\,\omega^2\,\z4
  +\omega^3\big(
  -256\,\z2\,\z3
  +360\,\z5\big)
\nonumber\\&
  +\omega^4\left(
  144\,\zeta_3^2
  +\z6\,\frac{58}{3}\right)
\nonumber\\&
  +\omega^5\big(
  -280\,\z3\,\z4
  -384\,\z2\,\z5
  +707\,\z7\big)
\nonumber\\&
  +\omega^6\left(
  272\,\z3\,\z5
  -\frac{144}{5}\zeta_{5,3}
  +\frac{1183}{30}\z8\right)
\nonumber\\&
  +\omega^7\big(
  -408\,\z4\,\z5
  -266\,\z3\,\z6
  -512\,\z2\,\z7
  +1185\,\z9\big)
\nonumber\\&
  +\omega^8\left(
  262\,\z3\,\z7
  -\frac{141}{7}\zeta_{7,3}
  +\frac{977}{7}\zeta_5^2
  +\frac{12097}{280}\zz10\right)
\nonumber\\&
  +\omega^9\left(
  -394\,\z5\,\z6
  -536\,\z4\,\z7
  -640\,\z2\,\z9
  -\frac{519}{2}\z3\,\z8
  +\frac{28677}{16}\zz11\right)
\nonumber\\&
  +\omega^{10}\left(
  298\,\z5\,\z7
  +258\,\z3\,\z9
  -\frac{46}{3}\zeta_{9,3}
  +\frac{1161445}{66336}\zz12\right)
\Bigg]
\nonumber\\&\hspace*{-10mm}
+g^6\Bigg[
  256\,\z4
  +\omega\big(
  1152\,\z2\,\z3
  -960\,\z5\big)
  +\omega^2\left(
  -1056\,\zeta_3^2
  -\frac{5000}{3}\z6\right)
\nonumber\\&
  +\omega^3\big(
  -1072\,\z3\,\z4
  +6880\,\z2\,\z5
  -6412\,\z7\big)
\nonumber\\&
  +\omega^4\left(
  3552\,\z2\,\zeta_3^2
  -6080\,\z3\,\z5
  +384\,\zeta_{5,3}
  -\frac{9170}{3}\z8\right)
\nonumber\\&
  +\omega^5\left(
  -1504\,\zeta_3^3
  -968\,\z4\,\z5
  +18648\,\z2\,\z7
  -\frac{187304}{9}\z9
  -\frac{20704}{3}\z3\,\z6\right)
\nonumber\\&
  +\omega^6\bigg(
  3856\,\zeta_3^2\,\z4
  +10976\,\z2\,\z3\,\z5
  -4496\,\zeta_5^2
  -576\,\z2\,\zeta_{5,3}
\nonumber\\&\qquad\qquad
  -10332\,\z3\,\z7
  +648\,\zeta_{7,3}
  -\frac{9363}{5}\zz10\bigg)
\nonumber\\&
  +\omega^7\bigg(
  256\,\z2\,\zeta_3^3
  -3968\,\zeta_3^2\,\z5
  -12944\,\z5\,\z6
  +1533\,\z4\,\z7
  -\frac{3095713}{60}\zz11
\nonumber\\&\qquad\qquad
  -\frac{615577}{45}\z3\,\z8
  +\frac{576}{5}\zeta_{3,5,3}
  +\frac{4608}{5}\z3\,\zeta_{5,3}
  +\frac{357092}{9}\z2\,\z9\bigg)
\nonumber\\&
  +\omega^8\bigg(
  -128\,\zeta_3^4
  +192\,\zeta_{4,4,2,2}
  +18472\,\z3\,\z4\,\z5
  +9348\,\z2\,\z3\,\z7
  -7294\,\z5\,\z7
\nonumber\\&\qquad\qquad
  -\frac{143216}{9}\z3\,\z9
  -\frac{5484}{7}\z2\,\zeta_{7,3}
  +\frac{2112}{5}\z4\,\zeta_{5,3}
  +\frac{2512}{3}\zeta_{9,3}
\nonumber\\&\qquad\qquad
  +\frac{4876}{3}\zeta_3^2\,\z6
  +\frac{39828}{7}\z2\,\zeta_5^2
  +\frac{449442803}{82920}\zz12\bigg)
\Bigg]
\nonumber\\&\hspace*{-10mm}
+g^8\Bigg[
  384\,\zeta_3^2
  +\frac{1888}{3}\z6
  +\omega\big(
  7232\,\z3\,\z4
  -20736\,\z2\,\z5
  +8848\,\z7\big)
\nonumber\\&
  +\omega^2\left(
  -20224\,\z2\,\zeta_3^2
  +24384\,\z3\,\z5
  -\frac{3968}{5}\zeta_{5,3}
  +\frac{1646704}{45}\z8\right)
\nonumber\\&
  +\omega^3\left(
  11712\,\zeta_3^3
  +84432\,\z4\,\z5
  +58256\,\z3\,\z6
  -165184\,\z2\,\z7
  +\frac{935056}{9}\z9\right)
\nonumber\\&
  +\omega^4\bigg(
  47344\,\zeta_3^2\,\z4
  -204224\,\z2\,\z3\,\z5
  +108768\,\z3\,\z7
  -\frac{32456}{7}\zeta_{7,3}
\nonumber\\&\qquad\qquad
  +\frac{16256}{5}\z2\,\zeta_{5,3}
  +\frac{307512}{7}\zeta_5^2
  +\frac{34103276}{525}\zz10\bigg)
\nonumber\\&
  +\omega^5\bigg(
  506717\,\zz11
  -50432\,\z2\,\zeta_3^3
  -2976\,\zeta_{3,5,3}
  +88848\,\zeta_3^2\,\z5
  +331188\,\z4\,\z7
\nonumber\\&\qquad\qquad
  -\frac{1966496}{3}\z2\,\z9
  -\frac{58592}{5}\z3\,\zeta_{53}
  +\frac{757764}{5}\z3\,\z8
  +\frac{765064}{3}\z5\,\z6\bigg)
\nonumber\\&
  +\omega^6\bigg(
  17784\,\zeta_3^4
  -2856\,\zeta_{4,4,2,2}
  +97216\,\z3\,\z4\,\z5
  -465784\,\z2\,\z3\,\z7
  -\frac{112996}{9}\zeta_{9,3}
\nonumber\\&\qquad\quad
  -\frac{1486152}{7}\z2\,\zeta_5^2
  -\frac{81288}{5}\z4\,\zeta_{5,3}
  +\frac{100344}{7}\z2\,\zeta_{7,3}
  +\frac{324910}{3}\z5\,\z7
\nonumber\\&\qquad\quad
  +\frac{599780}{3}\zeta_3^2\,\z6
  +\frac{2520308}{9}\z3\,\z9
  +\frac{37008793}{13820}\zz12\bigg)
\Bigg]
\nonumber\\&\hspace*{-10mm}
+g^{10}\Bigg[
  7168\,\z2\,\zeta_3^2
  -3072\,\z3\,\z5
  -\frac{1104928}{45}\z8
  +\frac{5376}{5}\zeta_{5,3}
\nonumber\\&
  +\omega\left(
  -9984\,\zeta_3^3
  -200896\,\z4\,\z5
  +335872\,\z2\,\z7
  -\frac{309248}{3}\z3\,\z6
  -\frac{216224}{3}\z9\right)
\nonumber\\&
  +\omega^2\bigg(
  -300736\,\zeta_3^2\,\z4
  +754688\,\z2\,\z3\,\z5
  -235680\,\z3\,\z7
  -\frac{357051032}{525}\zz10
\nonumber\\&\qquad\quad
  -\frac{752816}{7}\zeta_5^2
  +\frac{48128}{5}\z2\,\zeta_{5,3}
  +\frac{74608}{7}\zeta_{7,3}\bigg)
\nonumber\\&
  +\omega^3\bigg(
  323072\,\z2\,\zeta_3^3
  -390272\,\zeta_3^2\,\z5
  -2974304\,\z4\,\z7
  -\frac{7598556}{5}\zz11
  +\frac{129024}{5}\zeta_{3,5,3}
\nonumber\\&\qquad\quad
  -\frac{38284336}{45}\z3\,\z8
  -\frac{3105200}{3}\z5\,\z6
  +\frac{174848}{5}\z3\,\zeta_{5,3}
  +\frac{33289088}{9}\z2\,\z9\bigg)
\nonumber\\&
  +\omega^4\bigg(
  3968\,\z4422
  -4640256\,\z3\,\z4\,\z5
  -1293616\,\zeta_3^2\,\z6
  +5041344\,\z2\,\z3\,\z7
\nonumber\\&\qquad\quad
  -892496\,\z5\,\z7
  +65632\,\zeta_{9,3}
  -\frac{42084279598}{31095}\zz12
  -\frac{15822496}{9}\z3\,\z9
\nonumber\\&\qquad\quad
  -\frac{419840}{3}\zeta_3^4
  -\frac{292832}{7}\z2\,\zeta_{73}
  +\frac{33728}{5}\z4\,\zeta_{5,3}
  +\frac{15677344}{7}\z2\,\zeta_5^2\bigg)
\Bigg]
\nonumber\\&\hspace*{-10mm}
+g^{12}\Bigg[
  96768\,\zeta_3^2\,\z4
  -148480\,\z2\,\z3\,\z5
  +10240\,\z2\,\zeta_{5,3}
  +4096\,\z3\,\z7
\nonumber\\&\qquad\quad
  -\frac{52672}{7}\zeta_{7,3}
  -\frac{22720}{7}\zeta_5^2
  +\frac{5404288}{21}\zz10
\nonumber\\&
  +\omega\bigg(
  5058656\,\z4\,\z7
  -267264\,\z2\,\zeta_3^3
  -34176\,\zeta_{3,5,3}
  +131456\,\zeta_3^2\,\z5
  -\frac{15587072}{3}\z2\,\z9
\nonumber\\&\qquad\quad
  -\frac{216192}{5}\z3\,\zeta_{5,3}
  +\frac{1031216}{3}\zz11
  +\frac{2505344}{3}\z5\,\z6
  +\frac{11939072}{15}\z3\,\z8\bigg)
\nonumber\\&
  +\omega^2\bigg(
  17248\,\zeta_{4,4,2,2}
  +15663712\,\z3\,\z4\,\z5
  +2173824\,\zeta_3^2\,\z6
  -13296896\,\z2\,\z3\,\z7
\nonumber\\&\qquad\quad
  +1893440\,\z5\,\z7
  -130608\,\zeta_{9,3}
  -\frac{41665760}{7}\z2\,\zeta_5^2
  -\frac{388704}{7}\z2\,\zeta_{7,3}
\nonumber\\&\qquad\quad
  +\frac{563552}{3}\zeta_3^4
  +\frac{906336}{5}\z4\,\zeta_{5,3}
  +\frac{5426096}{3}\z3\,\z9
  +\frac{309691704914}{31095}\zz12\bigg)
\Bigg]
\nonumber\\&\hspace*{-10mm}
+g^{14}\Bigg[
  -26624\,\zeta_3^4
  +7680\,\z4422
  -2988544\,\z3\,\z4\,\z5
  +1220096\,\z2\,\z3\,\z7
\nonumber\\&\qquad\quad
  -\frac{47719021024}{31095}\zz12
  -\frac{823808}{3}\zeta_3^2\,\z6
  -\frac{746496}{7}\z2\,\zeta_{7,3}
  +\frac{457216}{3}\z3\,\z9
\nonumber\\&\qquad\quad
  +\frac{559360}{9}\zeta_{9,3}
  +\frac{771584}{5}\z4\,\zeta_{5,3}
  +\frac{1607936}{3}\z5\,\z7
  +\frac{4011008}{7}\z2\,\zeta_5^2
\Bigg]\,.
\end{align}

%
%

\end{document}